\documentclass[ApJ,twocolumn]{aastex63}
\usepackage{gensymb}

\newcommand{\iso}[2]{$^{#1}$#2}

\graphicspath{{./}{figures/}}

\newcommand{\msun}{\ensuremath{M_\sun}}

\newcommand{\nue}{\ensuremath{\nu_{e}}}
\newcommand{\nuebar}{\ensuremath{\bar \nu_e}}
\newcommand{\numt}{\ensuremath{\nu_{\mu\tau}}}
\newcommand{\numtbar}{\ensuremath{\bar \nu_{\mu\tau}}}
\newcommand{\numu}{\ensuremath{\nu_{\mu}}}
\newcommand{\nutau}{\ensuremath{\nu_{\tau}}}
\newcommand{\numubar}{\ensuremath{\bar \nu_{\mu}}}
\newcommand{\nutaubar}{\ensuremath{\bar \nu_{\tau}}}

\newcommand{\percent}[1]{$#1\,\%$}
\newcommand{\MPA}{Max-Planck Institute for Astrophysics, Postfach 1317, 85741 Garching, Germany}
\newcommand{\UTphys}{Department of Physics and Astronomy, University of Tennessee, Knoxville, TN 37996-1200, USA}
\newcommand{\ORNLphys}{Physics Division, Oak Ridge National Laboratory, P.O. Box 2008, Oak Ridge, TN 37831-6354, USA}

\newcommand{\NCCS}{National Center for Computational Sciences, Oak Ridge National Laboratory, P.O. Box 2008, Oak Ridge, TN 37831-6164, USA}
\newcommand{\FAU}{Department of Physics, Florida Atlantic University, 777 Glades Road, Boca Raton, FL 33431-0991, USA}

\newcommand{\chimera}{{\sc Chimera}}
\newcommand{\xnet}{{\sc XNet}}
\begin{document}

\title{Tracer particles for core-collapse supernova nucleosynthesis: The advantages of moving backward}

\correspondingauthor{Andre Sieverding}
\email{asieverd@mpa-garching.mpg.de}

\shorttitle{Tracer particles for CCSN nucleosynthesis}
\shortauthors{Sieverding et al.}

\author[0000-0001-8235-5910]{Andre Sieverding}
\affiliation{\MPA}
\affiliation{\ORNLphys}
\author[0000-0002-8613-3140]{Preston G. Waldrop}
\affiliation{\UTphys}
\author[0000-0003-3023-7140]{J. Austin Harris}
\affiliation{\NCCS}
\author[0000-0002-9481-9126]{W. Raphael Hix}
\affiliation{\ORNLphys}
\affiliation{\UTphys}
\author[0000-0002-5231-0532]{Eric J. Lentz}
\affiliation{\UTphys}
\affiliation{\ORNLphys}
\author[0000-0003-0999-5297]{Stephen W. Bruenn}
\affiliation{\FAU}
\author[0000-0002-5358-5415]{O. E. Bronson Messer}
\affiliation{\NCCS}
\affiliation{\ORNLphys}
\affiliation{\UTphys}

\begin{abstract}
   After decades, the theoretical study of core-collapse supernova explosions is moving from parameterized, spherically symmetric models to increasingly realistic multi-dimensional simulations. 
   However, obtaining nucleosynthesis yields based on such multi-dimensional core-collapse supernova (CCSN) simulations, is not straightforward and frequently tracer particles are employed. 
   Tracer particles may be tracked in situ during the simulation, but often they are reconstructed in a post-processing step based on the information saved during the hydrodynamics simulation. 
   Reconstruction can be done in a number of ways and here we compare the approaches of backward and forward integration of the equations of motion to the results based on inline particle trajectories. 
   We find that both methods agree reasonably well with the inline results for isotopes for which a large number of particles contribute.  
   However, for rarer isotopes that are produced only by a small number of particle trajectories, deviations can be large.  
   For our setup, we find that backward integration leads to a better agreement with the inline particles by more accurately reproducing the conditions following freeze-out from nuclear statistical equilibrium, because the establishment of nuclear statistical equilibrium erases the need for detailed trajectories at earlier times. 
   Based on our results, if inline tracers are unavailable, we recommend backward reconstruction to the point when nuclear statistical equilibrium last applied, with an interval between simulation snapshots of at most 1~ms for nucleosynthesis post-processing.
\end{abstract}


\section{Introduction}
\label{sec:intro}
Core-collapse supernovae (CCSNe) have long been identified as a crucial astrophysical scenario for understanding the origin of the elements \citep{Woosley.Weaver:1995,Thielemann.Nomoto.ea:1996}.  
Most of our knowledge about CCSN nucleosynthesis is currently still based on spherically symmetric models with parameterized explosions \citep{Woosley.Heger.ea:2002,Heger.Woosley:2010,Sukhbold.Ertl.ea:2016,SuYoSh18,Curtis.Ebinger.ea:2019}. 
Self-consistent, multi-dimensional models of CCSNe, which can capture the turbulent fluid motions in the neutrino-heated regions above the proto-neutron star, however, have emerged recently to encompass the extreme physical and computational complexity of the problem \citep[see ][ for reviews]{Janka:2012,Janka:2016,Mueller:2016,Burrows.Vartanyan:2021}.
These modern, multi-dimensional simulations still make strong compromises regarding the detailed composition, by carrying  very limited nuclear reaction networks or none at all. 
This makes conclusions about the detailed composition more difficult and typically requires a separate post-processing step involving tracer particles, i.e., passive mass elements 
that are advected according to the fluid velocity.
Many studies have employed the tracer particle method (TPM) to calculate nucleosynthesis yields for different scenarios. 
For example, \citet{Seitenzahl.Roepke.ea:2010} and \citet{Travaglio.Hillebrandt.ea:2004} calculated yield for models of thermonuclear supernovae, \citet{Reichert.Obergaulinger.ea:2022}  and \citet{Nishimura.Takiwaki.ea:2015} for magneto-rotational supernovae and \citet{Nagataki.Hashimoto.ea:1997,Wanajo.Mueller.ea:2018,Eichler.Nakamura.ea:2018,Sieverding.Mueller.ea:2020,Witt.Psaltis.ea:2021} for simulations of neutrino-driven CCSNe.

Ideally, tracer particles are evolved in situ, within the simulation, with time steps dictated by the hydrodynamic solvers, making the particle trajectories as accurate as the simulation allows. 
\citet{Harris.Hix.ea:2017} have pointed out some of the uncertainties that result from using such inline tracked tracer particles compared to direct evolution of the composition on the simulation grid with small reaction networks.

Such particles, however, are not available for every simulation, instead tracer particles are often reconstructed in an additional post-processing step, based on a number of simulation snapshots saved at intervals far longer than the hydrodynamic time step \citep[e.g.,][]{Witt.Psaltis.ea:2021,Wanajo.Mueller.ea:2018}. Several studies in the literature have addressed a number of issues that need to be considered when using tracer particles. For example,
\citet{Bovard.Rezzolla:2017} have studied the impact of the placement of tracer particles in simulations of binary neutron star (BNS) mergers. These authors have, however, focused on tracer particles that are followed inline during the simulation. Furthermore, the BNS merger simulations only cover very short times after the merger. 
CCSN simulations, in contrast, are now able to follow the evolution for several seconds, until most nucleosynthesis is completed.  \citet{Seitenzahl.Roepke.ea:2010} have also discussed the initial placement of tracer particles for thermonuclear supernovae.
\citet{Nishimura.Takiwaki.ea:2015} have discussed in detail the spatial and mass resolution of tracer particles that are reconstructed by integration forward in time for simulations of magneto-rotational supernovae and $r$-process nucleosynthesis, but without inline tracked particles for comparison.
\citet{WoJaMu17} have used reconstructed tracer particles with a focus on the long-term evolution and the morphology of the remnant with simulations running beyond shock-breakout, but based on a parameterized explosion engine. In their work, some of the caveats we find in the reconstruction may be avoided by starting the tracer particle evolution at 15~ms after core bounce, rather than in the pre-collapse stellar model.  
Recently, \citet{Reichert.Obergaulinger.ea:2022} briefly discussed the advantages of reconstructing tracer particles for nucleosynthesis by integrating the equations of motion backward in time. 
These authors, however, do not have inline tracked particles for comparison and are limited to a 1~ms interval of their simulation snapshots.

Many other studies have discussed the aspects mentioned above and others, but when tracer particles are generated in a post-processing step, it is usually impossible to evaluate any uncertainties that result from this step, because no inline tracked tracer particles exist. 
Here, we try to elucidate the weaknesses of tracer particle reconstruction for the first time, by extracting tracer particles from a simulation that has already tracked them in situ. 
This allows a one-to-one comparison to evaluate the degree to which reconstructed tracer particles are reliable and to look for differences in the nucleosynthesis yields that result from the reconstruction step. 
In this article, we further provide an in-depth comparison of forward and backward integration for generating particles, using the results of two different simulations of neutrino-driven CCSN explosions based on snapshots saved every 0.2~ms. 
This also allows us to test the efficacy of tracer particles generated from snapshots saved less frequently. 
In agreement with \citet{Reichert.Obergaulinger.ea:2022} we find that backward reconstruction, up to the last point in time where nuclear statistical equilibrium (NSE) applied, can provide tracer particle histories generally suitable for nucleosynthesis calculations and we recommend intervals between simulation snapshots that are shorter than 1~ms.

In Section~\ref{sec:setup}, we first summarize the underlying supernova models and the massive star progenitors, as well as our post-processing framework. 
In Section~\ref{sec:trajectories}, we focus on the comparison between final and initial positions of the tracer particles and in Section~\ref{sec:conditions} we compare the nucleosynthesis conditions, most importantly, the electron fraction $Y_e$. Eventually in Section~\ref{sec:nucleosynthesis}, we compare the final integrated nucleosynthesis yields, including a discussion of the simulation snapshot interval before we conclude in Section~\ref{sec:conclusions}.

\section{Models and setup}
\label{sec:setup}
In the following, we briefly introduce  the axisymmetric (2D) explosion simulations with the \chimera\ radiation hydrodynamics code, the two massive star progenitor models used for this study and the explosion outcomes, as well as the reaction network code \xnet.

\subsection{Simulation}
\label{sec:simulations}

\chimera\ is a state of the art code for  multidimensional, radiation-hydrodynamics studies of stellar core collapse.  There are five principal components: hydrodynamics, neutrino transport, self-gravity, a nuclear reaction network, and a nuclear equation of state (EoS).
Full details of \chimera\ are documented in \cite{BrBlHi20}.
Hydrodynamics is evolved via a dimensionally split, Lagrangian-plus-remap scheme with piecewise parabolic reconstruction \citep[PPMLR;][]{CoWo84} as implemented in VH1 \citep{HaBlLi12}, modified to include the consistent multi-fluid advection of \citet{PlMu99}.
\chimera\ employs a radially moving grid to maintain good resolution at the surface of the proto-neutron star and the neutrinospheres.
The self-gravity is computed using a multipole expansion \citep{MuSt95}, replacing the Newtonian monopole with a general relativistic (GR) monopole \citep[][Case~A]{MaDiJa06}.
The neutrino transport solver is an improved and updated version of the multi-group (spectral), flux-limited diffusion (MGFLD) implementation of \citet{Brue85}, which uses near-complete physics, solving for four neutrino species (\nue, \nuebar, $\numt=\{\numu,\nutau\}$, $\numtbar=\{\numubar,\nutaubar\}$) while allowing for neutrino-neutrino scattering, pair exchange and other opacities.
Evolution of the composition within \chimera\ is handled by the Equation of State for regions that obey Nuclear Statistical Equilibrium (NSE) and by the open-source reaction network code XNet\footnote{https://github.com/starkiller-astro/XNet} in regions where NSE does not apply.
As documented in \cite{BrBlHi20}, models in the D-series include refinements to the neutrino transport scheme in the presence of shocks, to the transition of matter into and out of nuclear statistical equilibrium, to the adaptive radial mesh, as well as other small improvements and corrections when compared to models from the earlier C-series and B-series.

Tracer particles in \chimera\ are laid out in rows representing equal mass, covering the outer 0.1 \msun\ of the iron core outward to the outer boundary of the simulation.  
The absence of tracers from the inner iron core is an effort to concentrate the available tracers in material likely to be ejected.
In 2D simulations, tracers within a row distributed uniformly in the cosine of the latitude, therefore representing equal volumes within the row and hence equal masses.
In 2D D-series models, rows consist of 60 tracers.  
The number of rows then determines the tracer mass resolution.  

\subsection{Progenitor models and explosion outcome}
\label{sec:progenitors}
\chimera\ model D9.6-sn160-2D is based on a metal-free, low-mass Fe-core progenitor\footnote{z9.6; A. Heger 2012, private communication} that is an extension of \citet{Heger.Woosley:2010} and that has been studied by several groups in 1D, 2D and 3D \citep{Lentz:2022,SaHiMe21,Stockinger.Janka.ea:2020,Mueller.Tauris.ea:2019,Radice.Burrows.ea:2017,Mueller.Janka.Marek:2013,Janka.Huedepohl.ea:2012}. We will refer to this model in the following as D9.6.
The progenitor for this model has a very steep density gradient above the Fe-core, leading to a very low compactness that even allows for successful, neutrino-driven explosion in spherical symmetry. In multi-D simulations, this model exhibits early explosions with rapid shock expansion and it reaches the asymptotic explosion energy very quickly. Due to its similarities with bare ONeMg cores, both in initial structure and nucleosynthesis, it has been characterised as an electron-capture supernova-like model.
The nucleosynthesis of the innermost ejecta has been studied in detail by \citet{Wanajo.Mueller.ea:2018} based on an axisymmetric (2D) simulation. 
The 2D \chimera\ simulation used for this study was performed with a large nuclear reaction network including 160~nuclear species and run until 0.72~s after core bounce when the shock is in the He~shell at almost 20,000~km.
The diagnostic explosion energy is found to be  $0.19 \times 10^{51}$~erg for model D9.6, which is modestly larger than that for the 3D \chimera\ version, $0.17 \times 10^{51}$~erg \citep{SaHiMe21}, which was not run as long. 
It is substantially more than the value of $0.6\times 10^{50}$~erg reported by \citet{Wanajo.Mueller.ea:2018} for their 2D model and more than the value, $0.85\times 10^{50}$, found by \citet{Stockinger.Janka.ea:2020} in their 3D model.
For the nucleosynthesis, we find qualitatively very similar results as \citet{Wanajo.Mueller.ea:2018}, including large production of several neutron-rich isotopes, such as \iso{48}{Ca}, \iso{60}{Ni} as well as trans-Fe elements up to \iso{90}{Zr}.
The production of significant quantities of neutron-rich (stable) nickel, larger than the production of \iso{56}{Ni}, means that once the \iso{56}{Ni} decays, the supernova and its remnant will be nickel-rich, again looking more like an electron capture supernova than a core-collapse supernova.
Due to the very fast shock expansion, there is very little accretion onto the proto-neutron star (PNS), leading to rather low neutrino-luminosities and low-energy spectra. Due to the rather weak neutrino heating at later times, the entropy of the later, neutrino-driven proton-rich ejecta does not exceed 40 $k_B$/baryon. Both these factors inhibit the $\nu \rm{p}$ process from operating efficiently. 
The proton-rich ejecta are therefore characterized mostly by an alpha-rich freeze-out with $Y_e>0.5$ \citep{Pruet.Woosley.ea:2005}, with little nucleosynthesis beyond \iso{64}{Zn}. 
The fast explosion also leads to a relatively spherical morphology despite the axisymmetry.
In model D9.6, 3540 tracer particles are included, 603 of which become part of the ejecta, corresponding to $0.03$~\msun, which is the amount of unbound mass on the grid at the end of the simulation. We count particles with positive total energy and positive radial velocity as part of the ejecta. Each particle represents $5\times 10^{-5}$~\msun\ 
of material. Our mass resolution corresponds to the average resolution that was found to be numerically converged by \citet{Nishimura.Takiwaki.ea:2015} and a value of $10^{-5}$~\msun\ was also found to be converged by \citet{Seitenzahl.Roepke.ea:2010}. The steep density gradient of this progenitor model, however, leads to a relatively low spatial resolution for the outer parts of the domain (see Figure~\ref{fig:D96_position_compare}), that could be improved by deviating from equal mass tracer particles as suggested in \citet{Seitenzahl.Roepke.ea:2010}.

Model D10.9-HW10, hereinafter referred to as D10.9,  represents a slightly more massive supernova than model D9.6, but also from a zero-metallicity progenitor from \citet{Heger.Woosley:2010}, which, to our knowledge, has not been used for explosion simulations before. This model is, in many respects, much more of a traditional core-collapse supernova than model D9.6. 
The progenitor of D10.9 has a much shallower density profile, leading to a later onset of the explosion, more accretion, a more massive PNS and stronger neutrino heating. 
This simulation only included a small nuclear reaction network including 14 $\alpha$ nuclei as well as neutrons, protons and \iso{56}{Fe}. The simulation was run until almost 2.2~s after bounce when the explosion shock reaches  the He-shell. The diagnostic energy reaches around 0.21~$\times 10^{51}$~erg, with an estimated gravitational binding energy of  $0.01\times 10^{51}$~erg of overburden left from the material mostly exterior to the simulation domain. In spite of the larger ejecta mass, the explosion is only slightly more energetic than the explosion of D9.6, indicating lower velocities. The explosion is also much more delayed in D10.9 and the shock and the ejecta develop a dipolar morphology commonly seen in axisymmetric simulations. 
A detailed account of the explosion will be published together with a larger set of axisymmetric simulation results. 
The ejecta are mostly proton-rich with $Y_e>0.5$, but again, there is no significant $\nu \rm{p}$ process operating. There is also a neutron-rich component with $Y_e$ down to values of 0.42, which allows for the $\alpha$ process to produce elements up to Zr \citep{Woosley.Hoffman.ea:1992}. 
In this model,  $0.18 M_\odot$ of material are unbound on the grid. 5293 out of a total of 19020 tracer particles represent $0.16 M_\odot$ of ejected material, more than five times as much as in D9.6. 
The difference between the ejecta mass represented by the tracer particles and the unbound mass on the simulation grid mostly results from material with positive total energy, but negative radial velocities. This material is not included in the set of ejecta particles because their thermodynamic trajectories cannot be extrapolated. The mass per particle is $3\times 10^{-5}M_\odot$ and thus the mass resolution is almost twice as high as in model D9.6. 
In model D10.9 we also find 817 particles, corresponding to 0.025~\msun, that are still ahead of the shock at the end of the simulation. These particles are identified as particles that change in radial position by less than \percent{1} during the whole calculation. Such particles are predominantly located in the direction perpendicular to the symmetry axis where the shock expands more slowly and most of the particles are formally still bound. We exclude these particles from the set of ejecta particles used for the nucleosynthesis calculations because we cannot extrapolate the trajectory. 

\subsection{Reaction network}
\label{sec:network}
The nucleosynthesis calculations are performed using XNet, including more than 5000 nuclear species and nuclear reactions from the JINA Reaclib database \citep{Cyburt.Amthor.ea:2010}. 
Neutrino reactions are only included for neutrons and protons, which is sufficient to capture effects of the $\nu \rm{p}$~process \citep{Froehlich.Martinez-Pinedo.ea:2006}, but we do not include the reactions for the classical $\nu$~process \citep{Woosley.Hartmann.ea:1990}. In the following we will use the terms mass fractions, $X_i$, and abundances, $Y_i$, which are connected as $X_i=A_i\, Y_i$, where $A_i$ is the mass number of a given isotope.  

\subsection{Reconstruction and extrapolation}
\label{sec:reconstruction}
The position $\vec{x}(t)$ of a tracer particle at time $t$ is determined by the equation of motion, $\frac{d\vec{x}}{dt}(\vec{x},t)=\vec{v}(\vec{x},t)$, where $\vec{v}(\vec{x},t)$ is the fluid velocity at the particle's position, which is obtained from the discrete simulation grid by linear interpolation.
For the particles evolved within the \chimera\ simulations, this equation is solved for each particle with forward integration using Euler's method at the end of each dimensionally split hydrodynamic sub-step. 
The velocity field in this case is updated for each timestep of the simulation, which are usually of the order of a few $10^{-7}$ seconds. 
These timesteps are much smaller than the usual zone crossing time in the fluid where the tracers are active.
The reconstructed tracer particles, on the other hand, are based on the simulation snapshots that are periodically saved to disk during the simulation. For this study, snapshots are taken every $0.2\,\mathrm{ms}$, giving a much lower time resolution than tracking particles inline, leading to larger numerical errors in the integration of the equation of motion.
Similar to the approach outlined in \citet{Wanajo.Mueller.ea:2018}, we divide the integration for each timestep into a number of $N_{\rm{sub}}$ substeps of equal length covering a time interval $\Delta t$ each. 
For each of these substeps, the velocity field is linearly interpolated in time and space to the position of the tracer particle. 
The division into substeps prevents particles from skipping grid cells in the case of high velocities and large variations of the velocity in space. 
The results presented here are based on calculations using $N_{\rm{sub}}=40$, which leads to an effective time step of 0.005~ms in the integration of the equations of motion. We show in Section~\ref{sec:backward_accuracy}, that this is sufficient to prevent particles from jumping across more than one grid cell per iteration. Using an adaptive time step, as used in \citet{Reichert.Obergaulinger.ea:2022}, might be computationally more efficient. For each substep advancing from step $n$ to $n+1$, the time integration uses the second-order accurate midpoint method, i.e., 
we first calculate the position of the particle after half a time step based on the initial velocities to $t_{n+1/2}=t_{n}+1/2 \Delta t$ as 
\begin{equation}
\label{eq:midpoint1}
 \vec{x}(t_{n+1/2}) = \vec{x}(t_n) + \frac{1}{2} \Delta t \cdot \vec{v} ( \vec{x}(t_n), t_n),
\end{equation}
and then calculate the actual movement based on the velocity field at the midpoint location as 
\begin{equation}
\label{eq:midpoint2}
 \vec{x}(t_{n+1})    = \vec{x}(t_n) + \Delta t \cdot \vec{v} (\vec{x} (t_{n+1/2}), t_{n+1/2} ) .
\end{equation}
We compared to calculations with  $N_{\rm{sub}}=10$ and did not find noticeable differences. We also performed calculations using Runge-Kutta 4th order time integration and did not find a significant impact on the results (see also discusstion in Section~\ref{sec:backward_accuracy}). We do not expect a major impact from the integration scheme, since the time interval between simulation snapshots is the primary limiting factor.
In this paper, we compare two different approaches to the reconstruction and the agreement between inline tracked and reconstructed particle trajectories. 

   \paragraph{Forward} Here we only consider the most straightforward approach, where tracer particles are placed in the initial, spherically symmetric pre-collapse configuration and evolved forward following the fluid motion calculated by the simulation.
    In the spherically symmetric progenitor model, the initial particle positions follow the density profile, such that each particle represents the same mass, in the same way as for the initial tracer particle setup in the simulation. 
    
\paragraph{Backward} Tracer particles can also be placed at the end of the simulation and the equations of motion are integrated backward in time, re-tracing the fluid velocities. Formally, this is achieved with $\Delta t <0$ in equations (\ref{eq:midpoint1}) and (\ref{eq:midpoint2}), such that $t_{n+1}<t_n$. The backward trajectories obtained in this way are then inverted and used as input for the reaction network calculations in the physical forward direction.
   
\begin{figure}
    \centering
    \includegraphics[width=0.99\linewidth]{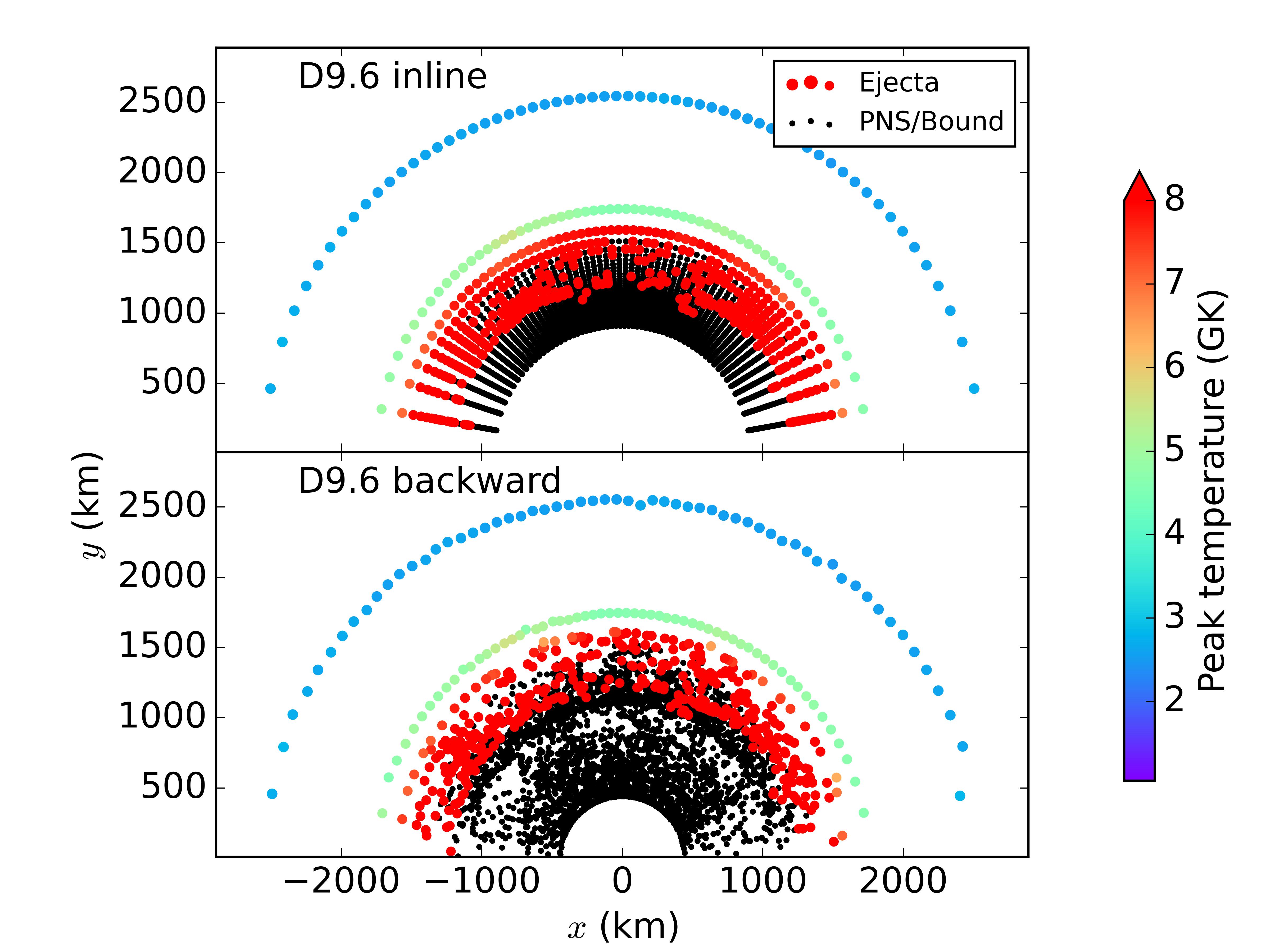}
\caption{Initial position of tracer particles in model D9.6 with the original layout in the top panel and the reconstructed position shown in the bottom panel. Colors indicate peak temperatures and black circles mark particles that are not part of the ejecta. The outer ray structure is well reproduced, while the inner tracer particles show significant differences.}
    \label{fig:D96_position_compare}
\end{figure}

In order to evaluate the accuracy of forward and backward integration, we take the inline tracer particles from the simulation as  fiducial reference results. For the forward reconstruction, we then consider tracer particles with the same initial locations as the original inline particles and integrate the equations of motion as described above based on the simulation snapshots.
For the backward reconstruction, we use the final positions of the inline particles and reconstruct trajectories by backward integration. In this way, we can assign the same mass to each particle in all cases and there is a clear one-to-one correspondence between particles across all the sets.  
In practice, for simulations without inline tracked particles, the placement of particles for backward integration needs to done differently (see, e.g. \citet{Wanajo.Mueller.ea:2018,Sieverding.Mueller.ea:2020,Reichert.Obergaulinger.ea:2022} for examples) and the assignment of the particle masses is less clear, potentially adding further uncertainties.  

Since our main interest is the calculation of nucleosynthesis yields with a large reaction network,
we are mostly interested in tracer particles that are eventually ejected by the explosion. We define the subset of ejected particles as those particle that achieve a positive total energy $E_{\rm{total},i}>0$  and positive radial velocity, $v_{\rm{radial},i}>0$ , at the end of the simulation.
For calculating the nucleosynthesis yields, we extrapolate the trajectories of the ejected particles in time with an initially exponential and later polynomial decrease of the density and adiabatic temperature, using the same approach as \citet{Harris.Hix.ea:2017}, to continue the reaction network calculations until 100~s after bounce. The local neutrino flux is extrapolated beyond the data from the simulation assuming an exponential decrease on a timescale of $\tau_\nu=3\,s$. Neutrinos are assumed to follow ideal Fermi-Dirac spectra with the average energies taken from the spectral neutrino transport of the simulation. The average energies are extrapolated assuming a linear decline reaching zero at 10 s after core bounce.

\section{Accuracy of reconstructed trajectories}
\label{sec:trajectories}
In addition to nucleosynthesis post-processing, tracer particles can be very useful to understand the history of the fluid flow, e.g., understanding where certain parts of the ejecta originate or the final fate of  material from a given part of the progenitor \citep{Bovard.Rezzolla:2017}. 
The details of the trajectory are also important, if corrections or modifications of the neutrino physics are studied \citep{Reichert.Obergaulinger.ea:2022}.
For nucleosynthesis, the origin of a particle is also particularly relevant, if temperature remains moderate and nuclear statistical equilibrium (NSE) is not established. In this case, the initial composition is also important (see Section~\ref{sec:non-nse}). 
In the following, the accuracy of the reconstruction of the initial position in the backward integration and the reproduction of the final fate with the forward reconstruction is discussed.

\begin{figure}
    \centering
    \includegraphics[width=0.99\linewidth]{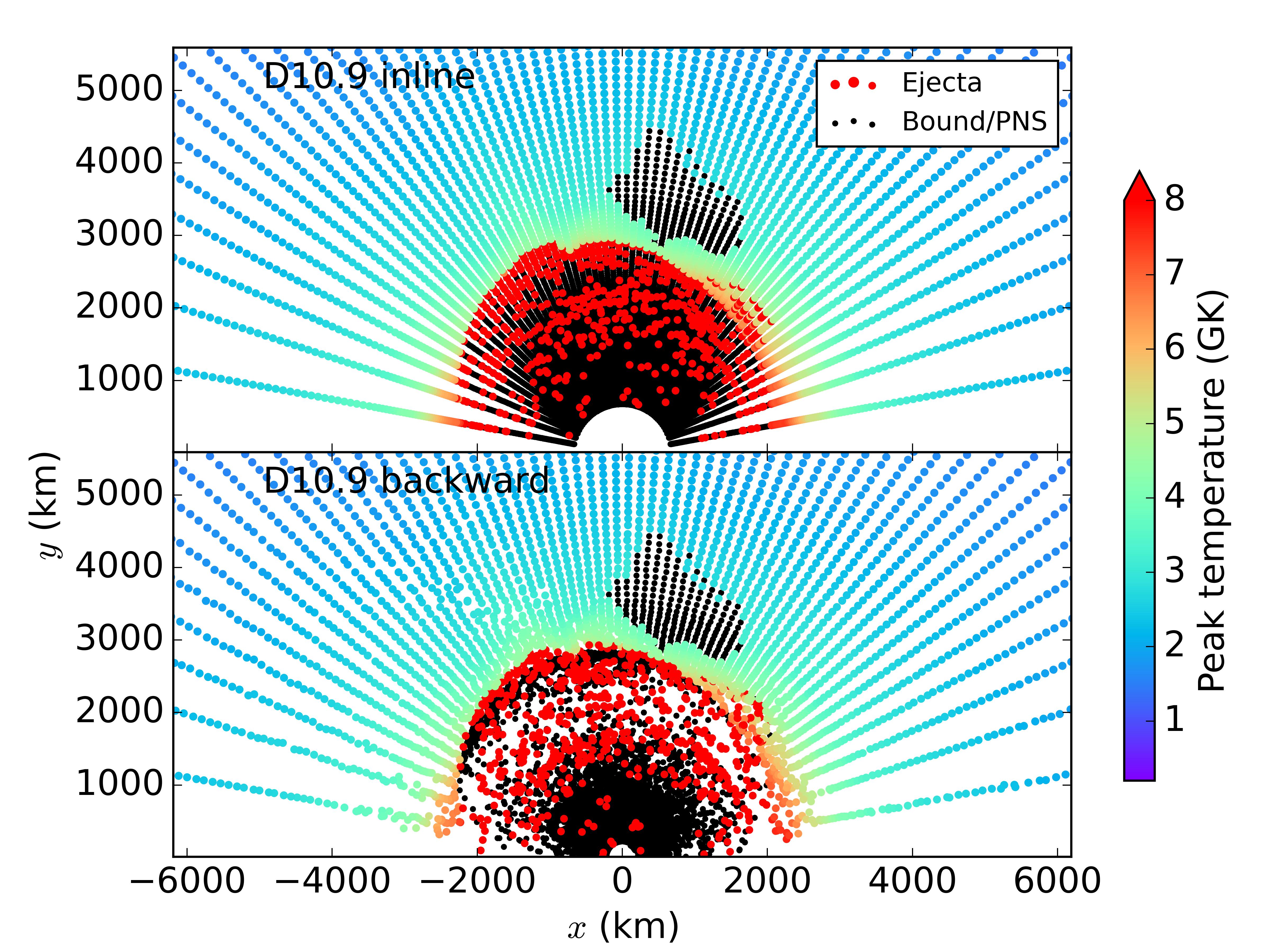}
\caption{Same as Figure \ref{fig:D96_position_compare} for model D10.9. The larger number of particles makes the reproduction of the ray structure more visible than in Figure \ref{fig:D96_position_compare}. }
    \label{fig:D109_position_compare}
\end{figure}

\subsection{Backward reconstruction of the initial positions}
\label{sec:backward_accuracy}
First, we compare the initial positions reconstructed by different methods of backward integration, starting from the final positions of our inline particles, to the original placement of the particles for the simulation.
The top panel of Figure~\ref{fig:D96_position_compare} shows the original layout of the tracer particles for model D9.6, which are positioned to represent equal mass as described in Section~\ref{sec:simulations}. 
The color code, corresponding to the peak temperature for ejected particles, in a useful proxy for the nucleosynthesis outcome.
Note that there is an additional row of particles at a radius of 6,500~km that is not shown to keep the inner particles visible. 
For comparison, the bottom panel of Figure~\ref{fig:D96_position_compare} shows the backward reconstruction of the initial positions of the tracer particles. 
Only the outer two rows of particles are well reproduced in Figure~\ref{fig:D96_position_compare} and this is also true for the outermost row that is not shown. 
For the inner particles, however, the original pattern is strongly distorted. 
The original rays are not recognizable but the radial range where the ejected particles with high peak temperature above 8~GK originate, between 1,200 and 1,800~km, is relatively well reproduced. 
It is also noticeable that, in the backward reconstruction, there are a large number of non-ejecta particles that originate from radii between 400 and 1,200~km, where there are no particles in the original layout. 

Figure \ref{fig:D109_position_compare} shows the comparison of the original layout and the backward reconstructed positions of the tracer particles from model D10.9. Here, we also do not show all of the outer particles for visibility. 
The structure of the rays in the original layout is well  reproduced by the reconstruction for most particles starting from radii larger than 2,500 km. 
The patch of bound accreted material, shown in black at an angle of around 80\degree (counter-clockwise) is also well reproduced.
There are, however, very noticeable differences for the innermost material, affecting mostly material that is not part of the ejecta. There is a clear gap in the positioning of the reconstructed particles around a radius of 2,000 km. 

\begin{figure}
  \centering
  \gridline{
\fig{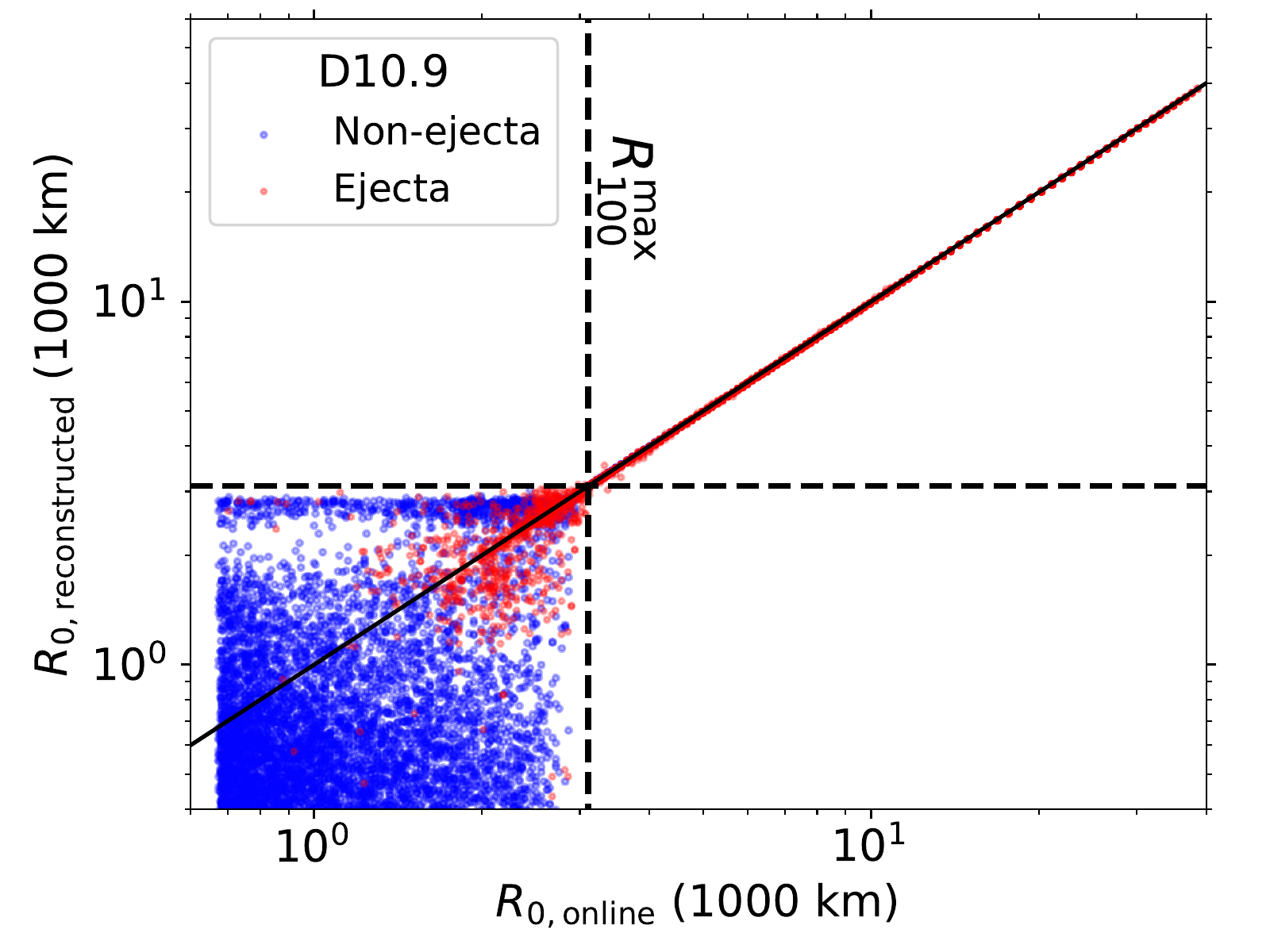}{0.99\linewidth}{(a)}}
\gridline{
\fig{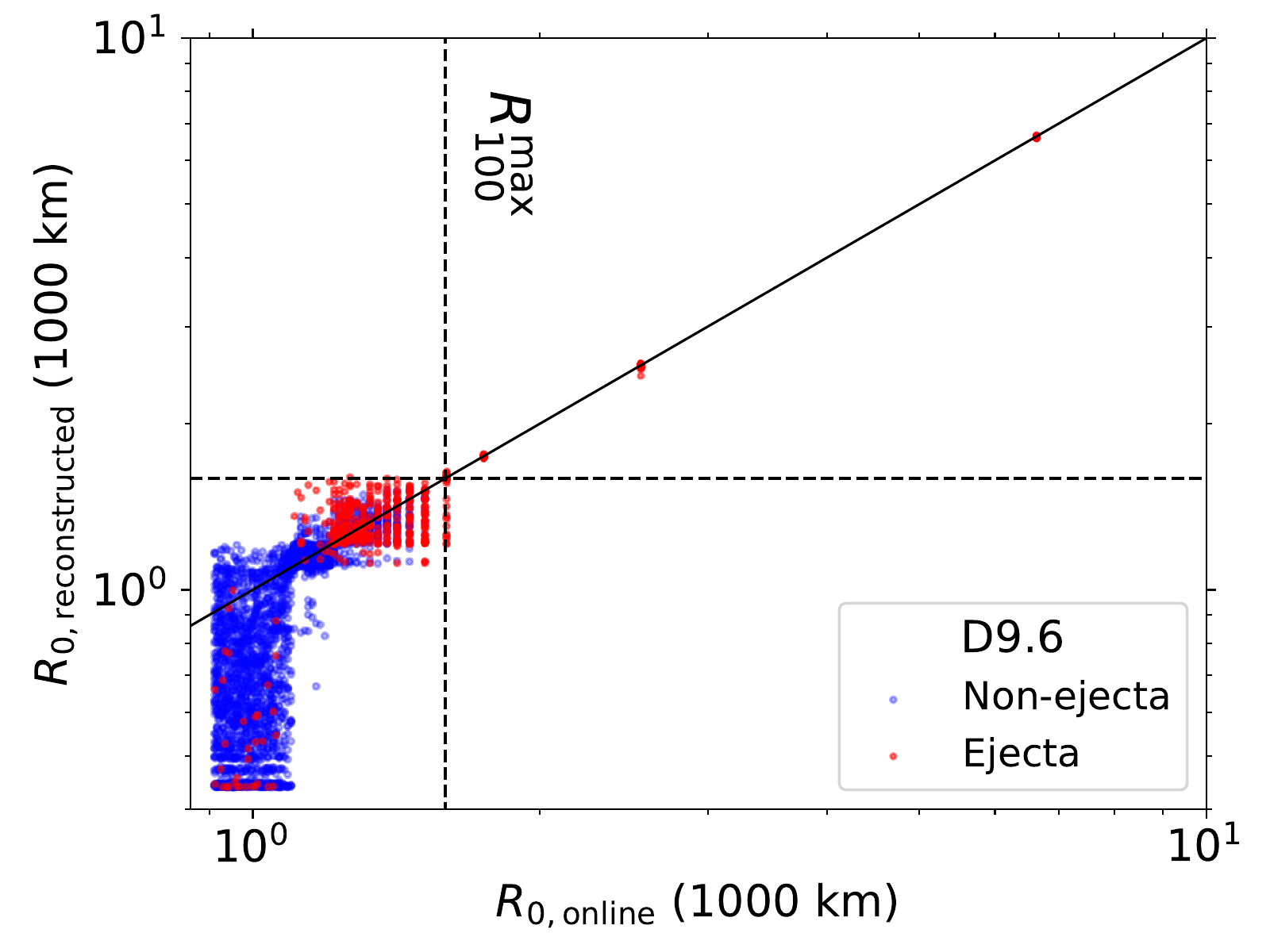}{0.99\linewidth}{(b)}}
\caption{Comparison of the initial position of the tracer particles from model D10.9 in the top panel (a) and from model D9.6 in the bottom panel (b). The x-axis shows the initial radius from the inline tracers while the y-axis shows the reconstructed initial radius. The vertical dotted line indicates $R^{\rm{max}}_{100}$ (see text), below which the differences between original and reconstructed positions increase drastically.}
\label{fig:initial_position_scatter}
\end{figure}

\begin{figure}
  \centering
  \gridline{
\fig{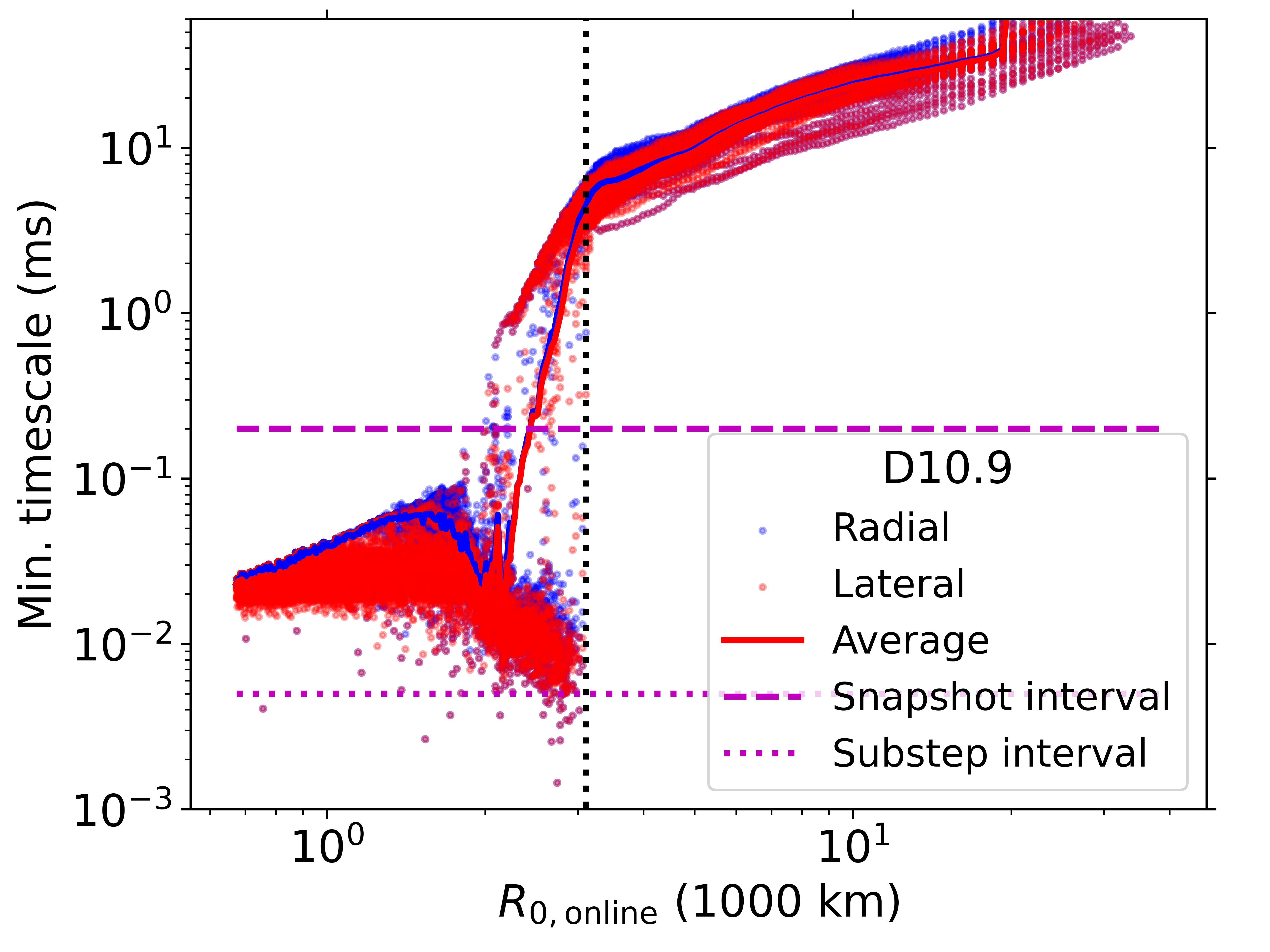}{0.99\linewidth}{(a)}}
\label{fig:D96_pf_non_nse}
\gridline{
\fig{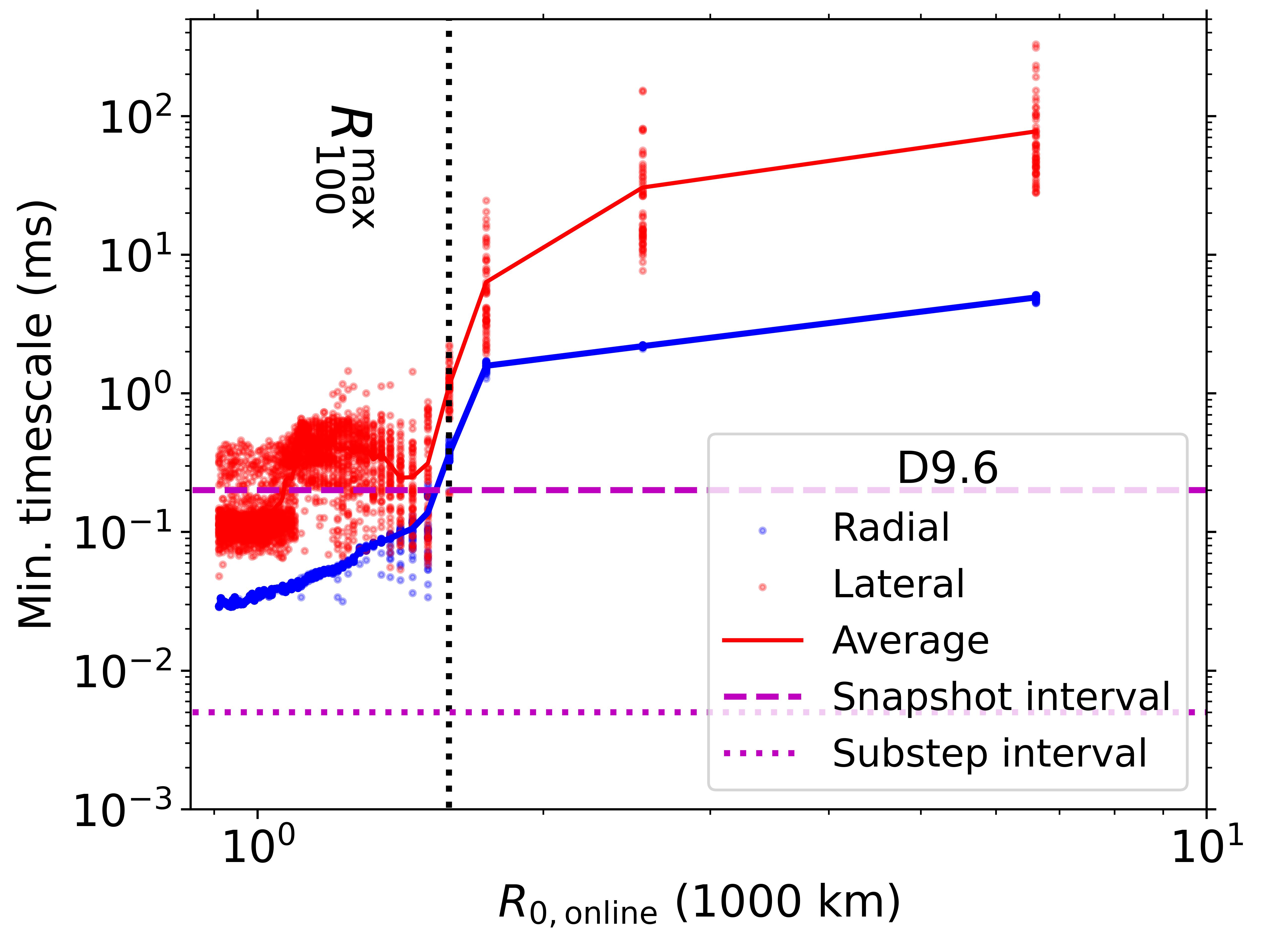}{0.99\linewidth}{(b)}}
\label{fig:D109_pf_non_nse}
\caption{Minimum timescale as defined in equations (\ref{eq:timescale_r}) and (\ref{eq:timescale_theta}) along the trajectory of each tracer particle as a function of the initial position from model D10.9 in the top panel (a) and from model D9.6 in the bottom panel (b). The average over each radial shell of tracer particles is indicated by solid lines. The vertical dotted line indicates $R^{\rm{max}}_{100}$ (see text) and the horizontal dashed line in indicates the 0.2~ms time interval between simulation snapshots, that effectively limits the backward reconstruction. The dotted magenta line indicates the effective time step for each substep of the integration with $N_{\rm{sub}}=40$. The lateral and radial components are very similar for model D10.9 (a) but show large differences for model D9.6 (b), which explodes more spherically.}
\label{fig:timescales}
\end{figure}

For a more direct comparison, Figure \ref{fig:initial_position_scatter}(a) shows the initial radial positions of the inline-tracked tracer particles,  $R_{0,\rm{inline}}$, compared to the radii that result from the backward reconstruction, $R_{0,\rm{reconstructed}}$, for model D10.9. 
The initial positions of particles above 3,000~km are well reproduced.
Between 2,000 and 3,000~km, the gap of reconstructed particles visible in Figure~\ref{fig:D109_position_compare} is also clearly visible in Figure~\ref{fig:initial_position_scatter}(a).
There is a very sharp transition between particles for which the initial position is well reproduced and particles for which deviations are large. 
In the following, we show that difficulties to reconstruct the initial position mostly affect particles that, at some point of their evolution, fall below a certain radius where they are involved in the complex convective motions around the neutrino heating region that are too fast to be followed accurately by the backward integration, even with simulation snapshots in 0.2~ms intervals. 

To illustrate this, we select the particles that fall below a radius of 100~km at some point during the simulation and collect their initial radial position into the set,
$\mathbb{R}_{0}({r_{\rm{min}}<100\,\rm{km}})$. 
The maximum value of this set, $R^{\rm{max}}_{100}$, is the radius in the progenitor model above which there are no particles that ever fall below $100\,$km. For model D10.9 the value of  $R^{\rm{max}}_{100}$ is 3,108~km, which is also indicated in Figure~\ref{fig:initial_position_scatter}(a) and agrees very well with the transition between relatively good reproduction of the initial position and the area with large deviations. The maximum value is not very dependent on the choice of 100~km as a threshold. $R^{\rm{max}}_{20}$, for example is 2,931~km, indicating a relatively sharp transition between the particles that get close to the PNS and are heated and those that are merely pushed out by the explosion. 
While this transition also roughly reflects the boundary between ejecta and accreted material, \percent{4.3} of the particles in $\mathbb{R}_{0}({r_{\rm{min}}<100\,\rm{km}})$ are part of the ejecta. Figure~\ref{fig:initial_position_scatter}(a) also illustrates that there are ejected particles with initial radii between 1,000 and 3,000 km. This also emphasizes that there is not a clear "mass-cut" in multi-dimensional supernova models. 

Figure \ref{fig:initial_position_scatter}(b) shows the comparison of the original and reconstructed initial positions for model D9.6. It is qualitatively very similar to model D10.9, with a sharp transition from a region of large differences for the inner particles to good agreement for the outer particles. In this model, the sampling of the outer tracer particles is very sparse because of the steep density gradient of the progenitor model. 
Only the outer three rows of particles are well reproduced. Due to the compact profile of the progenitor of D9.6, \percent{88} of all particles are initially below $R^{\rm{max}}_{100}=1,592\,\rm{km}$ and \percent{7} of the particles in $\mathbb{R}_{0}({r_{\rm{min}}<100\,\rm{km}})$ are part of the ejecta, corresponding to \percent{38} of the total ejecta.  Again, for particles with initial radii larger than this value, the reconstruction agrees very well with the original placement. Below $R^{\rm{max}}_{100}$, there are two distinct groups of particles. The outer group consists mainly of ejecta particles with a moderate degree of differences in the reconstructed initial radius that does not exceed a factor two. This is related to the unusual explosion mechanism of model D9.6, which does not depend on a large amount of neutrino heating. The outer group of particles is ejected within the first 300~ms after bounce, nearly directly when the explosion shock reaches the particles. The ejecta particles of the inner group, however, require convective neutrino heating to reach positive total energy and are ejected later, mostly after 500~ms.
This confirms that tracer particles representing the innermost ejecta, that are directly subject to neutrino heating and convective overturns, are the most likely to be inaccurate when a backward reconstruction method is used. This transition is sharper than the "mass cut" that divides ejecta from non-ejecta.

In order to further show that the failure to reconstruct the initial position is a consequence of the large time intervals between simulation snapshots, we calculate a timescale analogous to the numerical Courant condition, that requires that information should not cross more than one grid cell during one time step. We define a radial and lateral crossing timescale based on the velocities $v_{\rm{radial,i}}$ and  $v_{\rm{lateral,i}}$ for each particle $i$ as:
\begin{equation}
\tau_{\rm{C,radial,i}}=\Delta r(r_i(t),\theta_i(t))/v_{\rm{radial,i}}(t),
\label{eq:timescale_r}
\end{equation}
\begin{equation}
\tau_{\rm{C,lateral,i}}=\Delta \theta (r_i(t),\theta_i(t))/v_{\rm{lateral,i}}(t),
\label{eq:timescale_theta}
\end{equation}
where $\Delta r$ and $\Delta \theta$ are the size of the simulation grid cell in radial and lateral directions in which the particle is located at a given time $t$. 
For model D10.9, Figure \ref{fig:timescales}(a) shows the minimum values of  $\tau_{\rm{C,radial}}$ and $\tau_{\rm{C,lateral}}$ obtained along each particle trajectory as 
a function of the initial position of each particle in the original layout. The timescales are very short for the innermost particles, indicating high-velocity motions that cannot be adequately captured by the long intervals between simulation snapshots.
Note that we use an effective time step when reconstructing the tracer particle trajectories by dividing each time step into  $N_{\rm{sub}}=40$ substeps.  If particles cross more than one cell in one time step, not all of the information available from the simulation would be used, i.e., the information of the velocity field of the skipped cells, would be missed. The length of the substeps is also indicated in Figure \ref{fig:timescales}(a) and 
shows that the timescales are, with very few exceptions, always longer than our effective time step of the integration. This shows that the particles do not jump across several grid cells in one integration step. 

There is a steep increase in both timescales for particles with initial radii between $\approx 2,000$ km and where the average across radial shells crosses the value of 0.2~ms, which is the time interval between simulation snapshots, at around 2,500~km.  Above $R^{\rm{max}}_{100}$ at 3,108~km, most particles are not involved in motions with timescales that are shorter than the simulation snapshots. 
Figure \ref{fig:timescales}(b) shows the timescales for model D9.6. The explosion is more spherical than in D10.9 and neutrino heating plays a less important role for the onset of the explosion. Hence, the lateral timescales, indicative of convective turnovers, are generally longer than the timescales of the radial motion, which is shock-driven. Among the particles below $R^{\rm{max}}_{100}$, Figure \ref{fig:timescales}(b) also shows an outer group, for which the lateral velocities are close to, but mostly above, the snapshot interval and an inner group, for which both timescales are clearly shorter than the snapshot interval. The inner group corresponds to the group of particles with the largest deviations of reconstructed the initial position.

These cell crossing timescales, $\tau_{\rm{C,radial,i}}$ and $\tau_{\rm{C,lateral,i}}$,  can thus be used as criteria to determine whether reconstructed particles are suitable to make reliable connections between the position of material in the ejecta and the origin in the stellar model.
From Figure~\ref{fig:timescales}(a), we can also expect that an increase of the snapshot frequency by more than a factor 10 is required to safely capture the motion of all the particles, which is not practical.

The results in Figure \ref{fig:initial_position_scatter}(a) and Figure~\ref{fig:initial_position_scatter}(b) are largely independent of the integration method. We find very similar results using Runge-Kutta 4th order integration and with the midpoint method with less $N_{\rm{sub}}=10$. This indicates that the interval between simulation snapshots is the most important limiting factor, which higher-order integration methods have limited ability to cure.

\subsection{Initial composition}
\label{sec:non-nse}

\begin{figure}
  \centering
  \gridline{
\fig{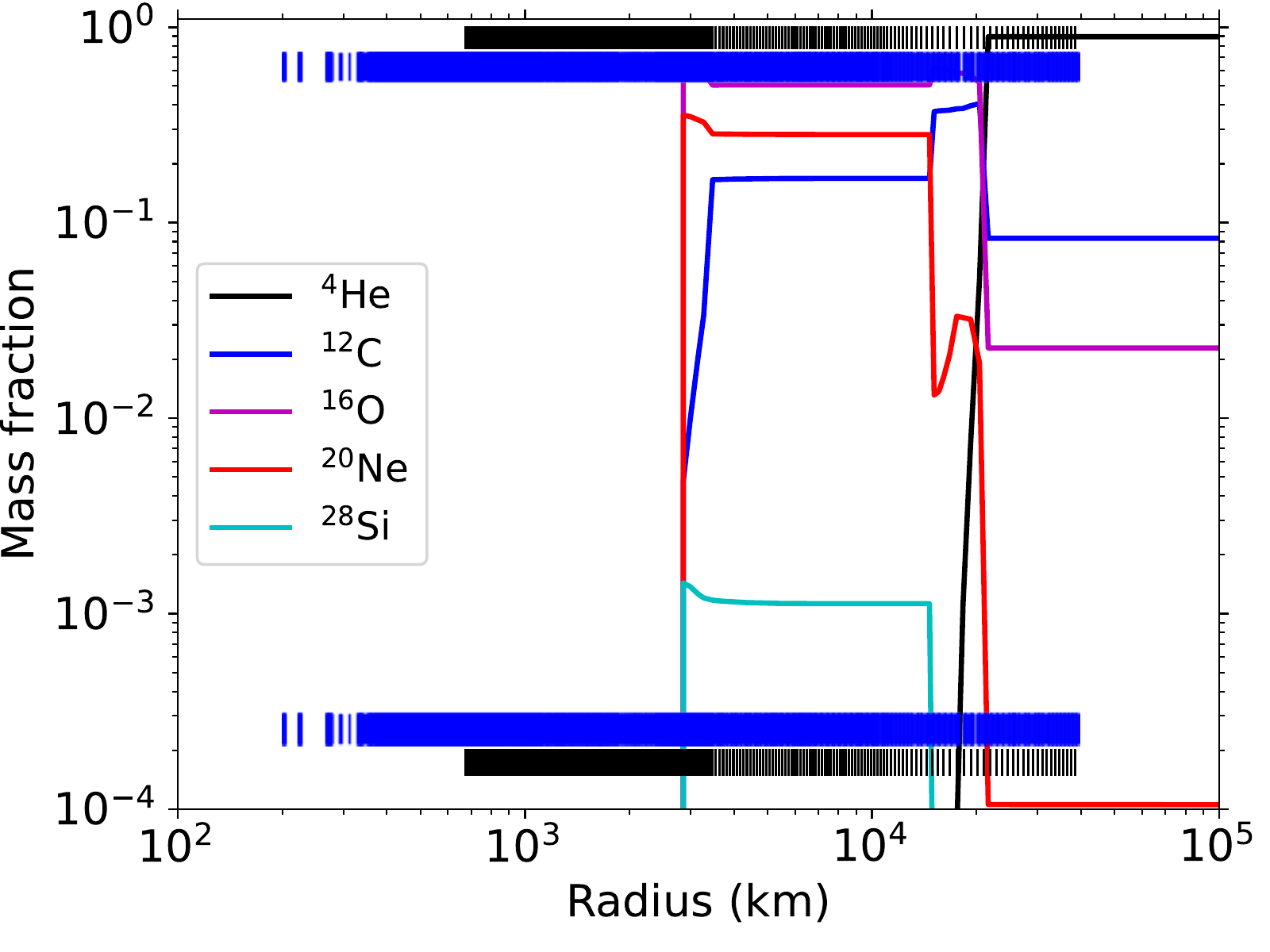}{0.99\linewidth}{(a) Model D10.9}}
\gridline{
\fig{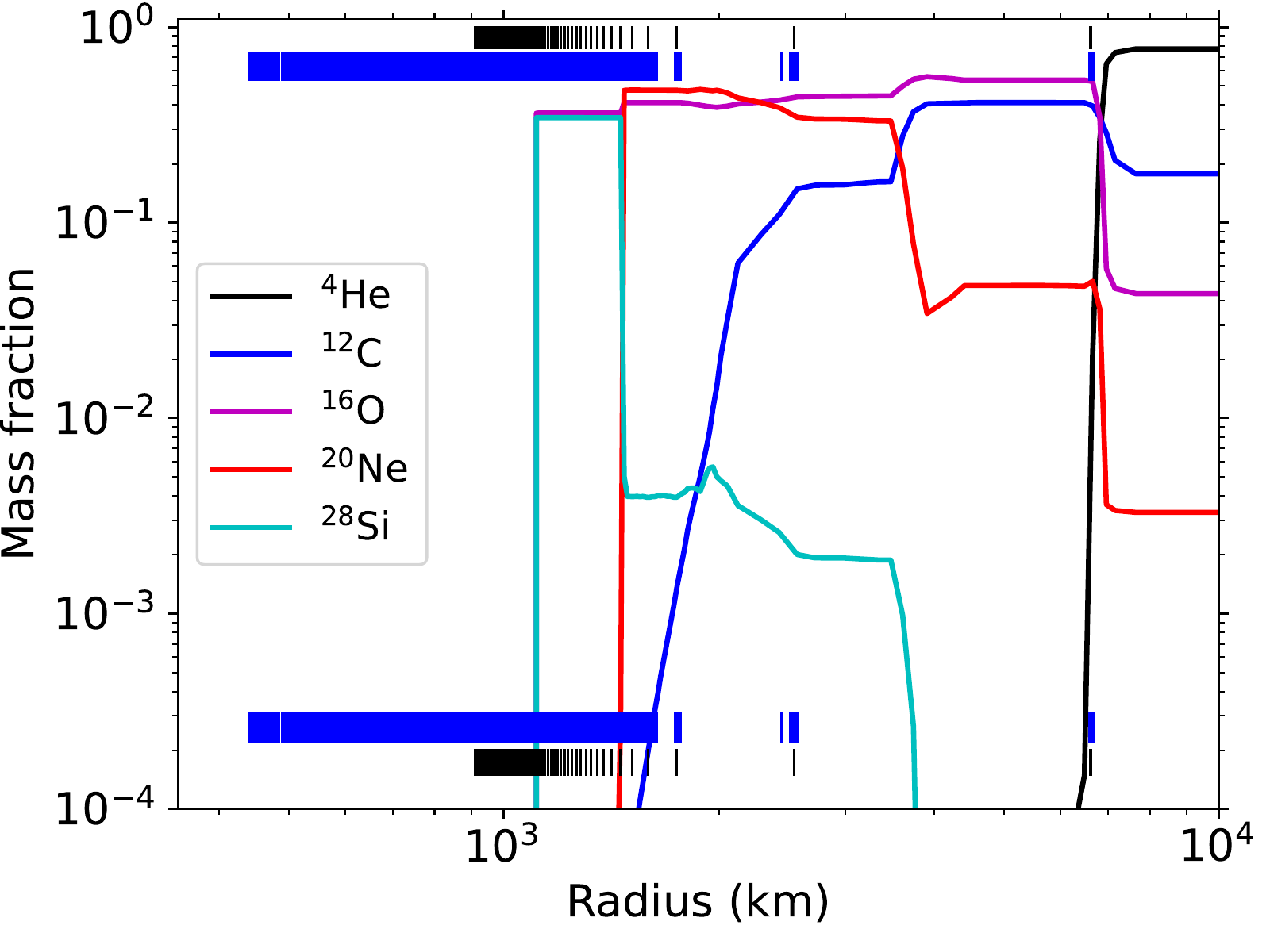}{0.99\linewidth}{(b) Model D9.6}}
\caption{Mass fractions of the most important isotopes for the progenitor of model D10.9 at the top (a) and for model D9.6 at the bottom (b). The vertical dashes at the top and at the bottom indicate the original tracer particle position (black) and the backward reconstructed 
position (blue). While the original layout is in strict shells, the reconstruction leads to a distribution around the original radii, as well as locations deeper in the Fe-core. For model D9.6 in (b),note that the outermost particles are spread across the transition to the He-shell, where large gradients in the composition are present.}
\label{fig:comp_and_loc}
\end{figure}

For nuclear reaction network calculations, the initial position is only important for particles that never reach conditions of NSE. 
For particle trajectories in which NSE is realized, the initial composition becomes irrelevant for the final nucleosynthesis as long as the electron fraction, $Y_e$, at freeze-out remains accurate. 
Note that the whole trajectory also becomes important if the $Y_e$-history needs to be reprocessed or corrected \citep[e.g.,][]{Sieverding.Mueller.ea:2020}.
For non-NSE particles nuclear reactions build on the initial composition taken from the pre-SN model based on the particles initial position. Hence, an inaccurate reconstruction of the initial position may impact the final nucleosynthesis result.
For the D9.6 model, we find 225 particles, among the 603 inline particles that are ejected, do not reach a temperature of more than 8~GK, which we take as the threshold to assume NSE, rather than evolving the reaction network throughout. 
These particles are traced to initial positions between 1,500~km and 7,000~km, corresponding to the three outermost shells of particles. 
The final total energy and velocity, and thus the classification as ejecta are the same for the backward reconstructed particles, but the peak temperatures can differ. There are 8 particles that reach more than 8~GK with the reconstruction that are non-NSE particles in the inline tracking and there are 3 non-NSE particles in the reconstructed set, that reach NSE in the inline tracking, giving only 220 non-NSE particles among the reconstructed particles.  
Among the particles that do not reach NSE in both the inline tracking and the reconstruction, the maximum difference between the position of the original particle and the reconstructed particle is 335~km with only six particles showing differences of more than 100~km. The average difference is 15~km with \percent{95} of non-NSE particles having a difference of less than 28~km. 
The initial radial position of the original and reconstructed particles are indicated by the short vertical dashes in Figure \ref{fig:comp_and_loc}(b).

\begin{figure}
  \centering
  \gridline{
\fig{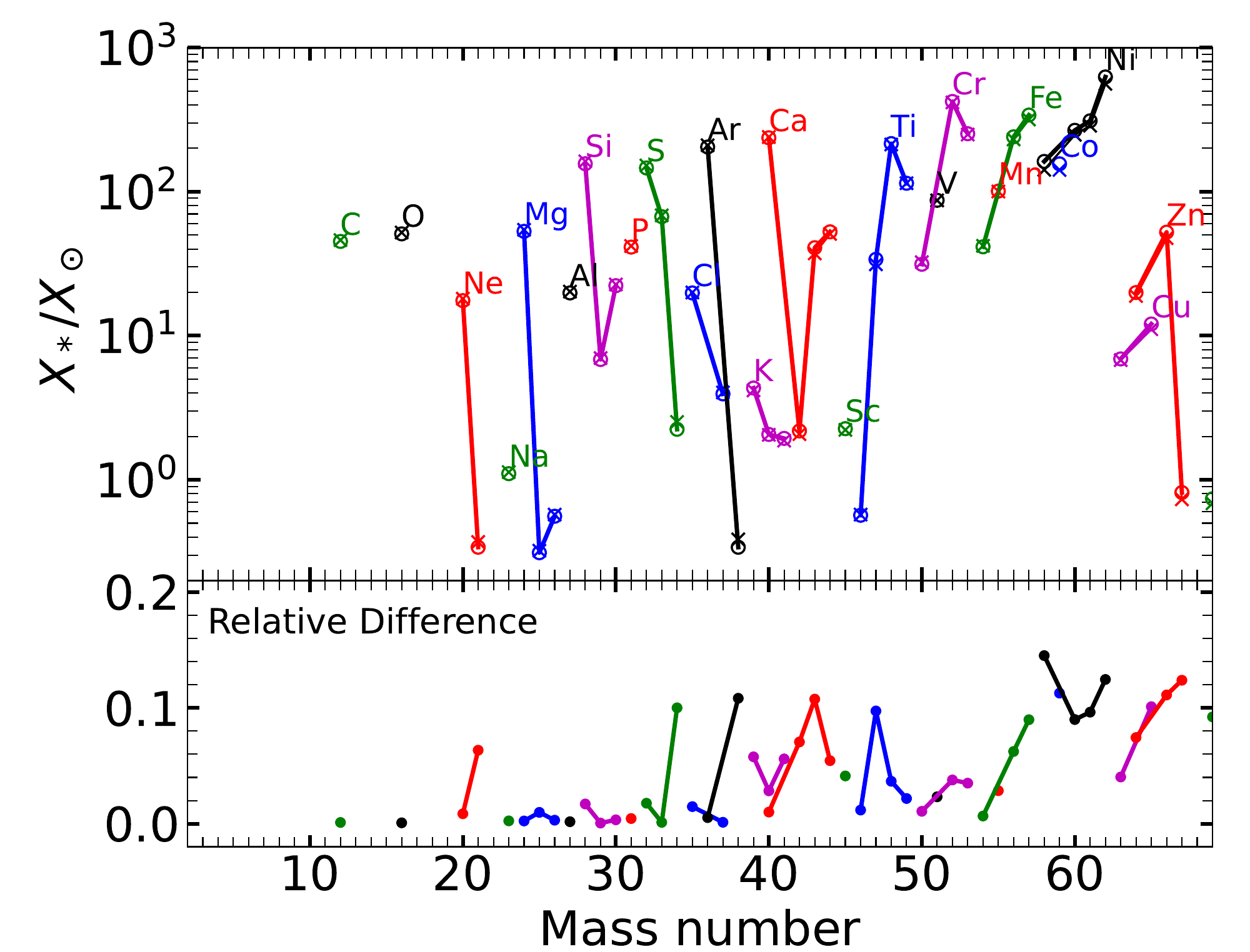}{0.99\linewidth}{(a)}}
\label{fig:D96_pf_non_nse}
\gridline{
\fig{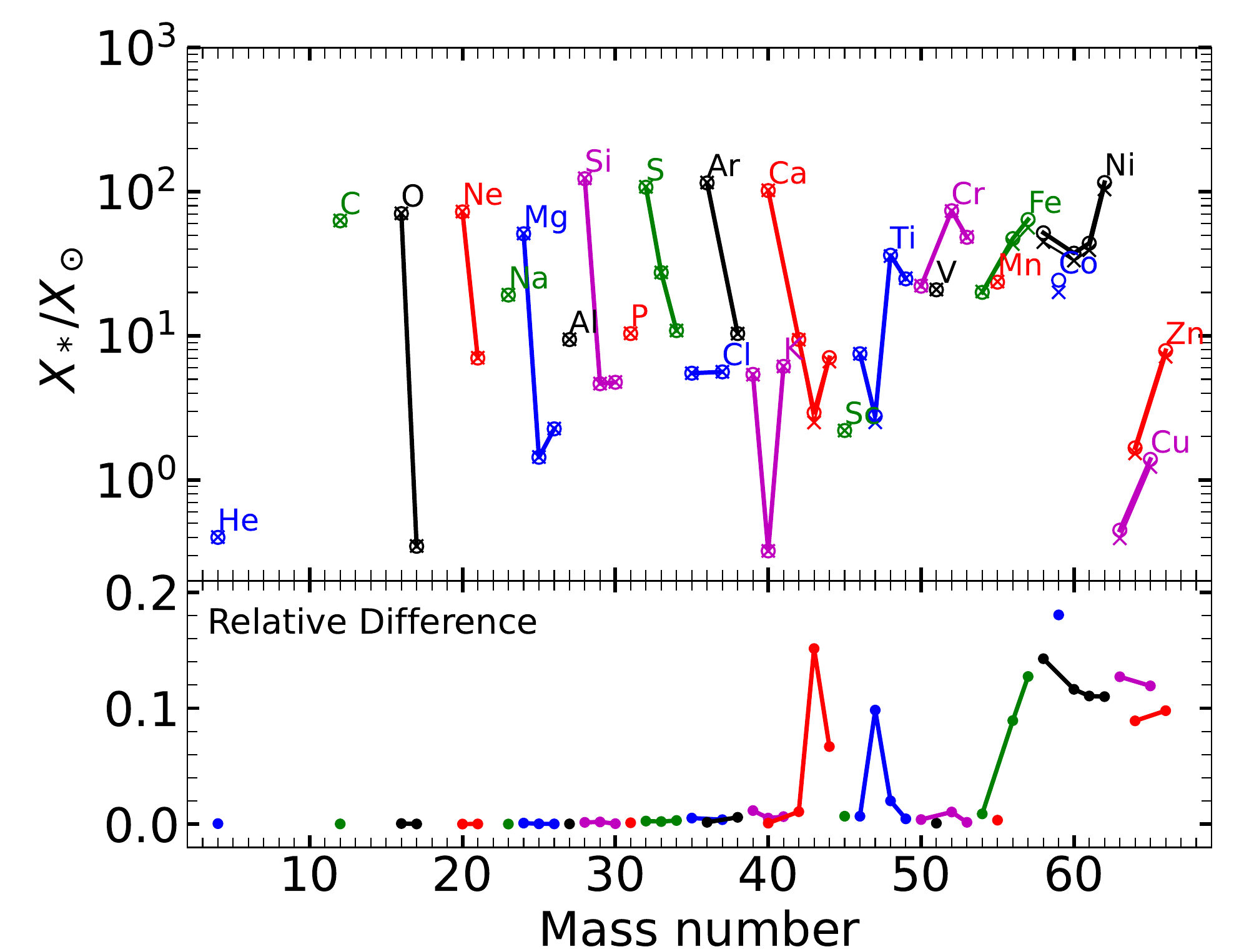}{0.99\linewidth}{(b)}}
\label{fig:D109_pf_non_nse}
\caption{Production factors (mass fractions relative to solar abundances) including only particles that do reach NSE for both, the inline and backward reconstructed trajectories. Model D9.6 is shown in the upper panel (a) and model D10.9 in (b). The lower panels show the relative difference when compared to the results with the backward reconstructed particles. The largest differences are found for isotopes produced in Si-burning, at temperatures close to NSE.}
\label{fig:pf_non_nse}
\end{figure}

Despite these small differences in initial radius, the initial location of the innermost non-NSE particles is on a steep compositional gradient at the interface of the O/Ne to the C shell, which leads to large differences in the initial composition. 
Considering only abundances larger than $10^{-4}$, we find 54 (\percent{25}) of the non-NSE particles have relative differences in the initial abundances compared to the inline tracers that are larger than \percent{10}. 
The majority those particles, however, experience peak temperatures exceeding 6~GK, and thus are close to NSE. Hence, the impact on the final composition is small. The results of the nucleosynthesis calculations including only tracers that do not reach temperatures exceeding 8~GK in both inline tracking and backward reconstruction are shown in Figure \ref{fig:pf_non_nse}. The top panel shows the production factors while the lower panel shows the relative difference. Differences are less than \percent{20} with the largest deviations for species that are produced in explosive Si- and O-burning, such as Ni, Fe, Ca, S and Ar~isotopes. Since these isotopes require high temperatures, they are affected more strongly by the peak temperatures and thermodynamic evolution rather than the initial composition.

In model D10.9, 5630 of the 6236 ejected inline tracer particles do not reach NSE, which is a much larger fraction than in model D9.6 because of the shallower density gradient of model D10.9, that covers more mass in the simulation domain. 
As a result, the non-NSE nucleosynthesis is much better sampled and the integrated outcome less sensitive to individual particles. Figure \ref{fig:comp_and_loc}(a) indicates the position of the tracer particles in the context of the progenitor's composition. The interfaces between different compositional shells are well resolved already with the original particles that are indicated by the black horizontal dashes. The backward integration leads to an increased spread of the radial position, affecting the innermost particles most strongly.
Regarding the initial radius of the non-NSE particles, however, there are only six particles with large differences of the initial position of more than 1,000~km, all of which are among the innermost non-NSE particles. The average difference, however, is only 13~km corresponding to less than \percent{1} of relative difference. For \percent{95} of non-NSE particles, the initial position is reproduced with a difference of less than 35~km. 
Considering only abundances larger than $10^{-4}$, we find only 285, or about \percent{5}, of the non-NSE particles have relative differences in the initial abundances that are larger than \percent{10}, which is a much better agreement than for model D9.6. This is mostly because of better sampling and an initial positioning of the tracer particles that is less aligned with compositional interfaces. While the outer two shells of Lagrangian particles represent two different compositional shells in model D9.6, there are several shells of particles for each compositional shell in model D10.9, making the results more robust. 
Figure \ref{fig:pf_non_nse}(b) shows the comparison of the nucleosynthesis results for model D10.9. The differences are very similar to the case of model D9.6, showing that the differences in the initial composition due to the tracer particle reconstruction affects the non-NSE contribution to the final yields by less than \percent{20}.
Overall, the particles that tend to have significant differences in the backward reconstructed initial position, tend to be particles that reach relatively high temperatures, reducing the sensitivity to the initial composition. 

\subsection{Final fate in the case of forward reconstruction}
\label{sec:final_fate}
In the forward reconstruction, the initial position is identical to the inline tracers by construction, but the final position at the end of the simulation and, thus, the fate of the particle, i.e., whether the particle will become part of the ejecta or not, can differ due to the reconstruction. Similar to the previous discussion of the initial position, we examine here the differences in the final fate for the forward reconstruction.
For model D9.6 in the forward reconstructed set, the final count of ejected particles is 598, making the ejecta mass agree within less than \percent{1}. While the total number is very similar, there are more than just 5 particles that are different. In total, there are 60 particles in the forward tracer particle set that are not ejected in the inline set and there are 65 particles that end up as ejecta in the inline set that are not part of the ejecta in the forward reconstruction, giving a total of 125 particles with a different prediction of the final fate. While this is just \percent{3.5} of all particles, it corresponds to \percent{20} of the ejected particles.
Among the particles that show a different fate, the largest initial radius is 1,512~km, corresponding to the particles in the pre-SN Si-core. Outside of the Si-shell, the fate of particles does not get changed by the reconstruction. Similar to the backward reconstruction, particles that experience the complex motions of the neutrino heated convection and fall deep toward the PNS are the most prone to show differences.

For model D10.9, we find 5,326 ejected particles in the forward reconstruction compared to 5293 based on the inline tracking, leading to less than \percent{1} difference in the ejecta mass. In total, however, the final fate of 821 particles differs, corresponding to about \percent{15} of the ejecta. 

\begin{figure}
  \includegraphics[width=\linewidth]{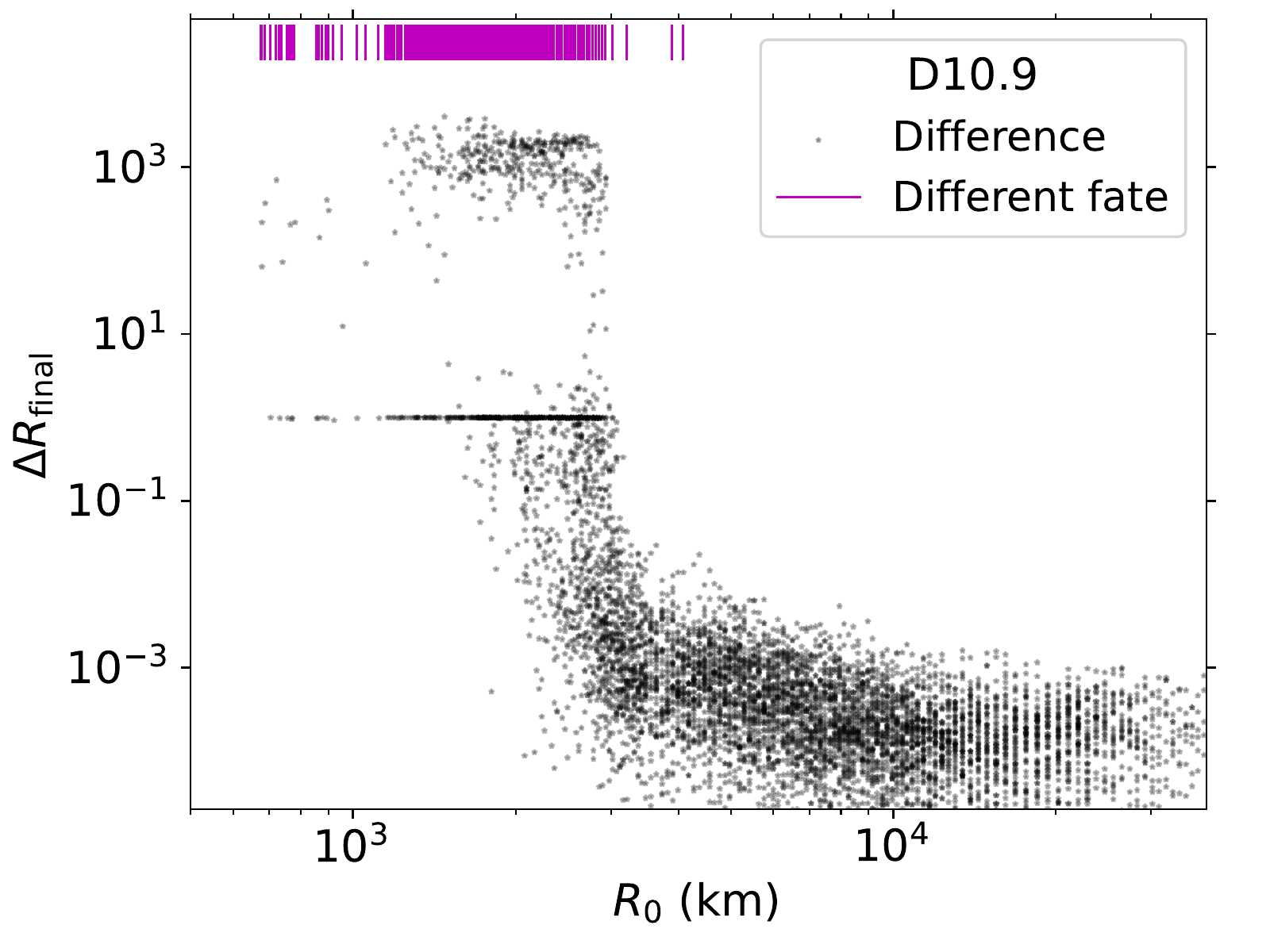}
  \caption{Relative difference of the final radius as a function of initial radius for model D10.9 with the inline tracking and the forward reconstruction. Vertical magenta lines at the top indicate positions of particles for which the final fate is different in the reconstruction compared to the inline tracking.}
  \label{fig:final_radius}
\end{figure}

To compare the forward reconstruction to the inline tracked particles, we compute the relative difference 
\begin{equation}
    \Delta R_{\rm{final}} = \frac{| R_{\rm{final,inline}}-R_{\rm{final,reconstr.}}|}{R_{\rm{final,inline}}}
\end{equation}
for each tracer particle.
Figure \ref{fig:final_radius} shows $\Delta R_{\rm{final}}$ as a function of the initial radius for all particles of model D10.9 that are part of the ejecta either in the inline tracked set or the reconstructed set. Particles that are not ejected in both cases are not shown. For forward reconstruction and the inline tracking, the initial radius is identical by construction.
The largest differences are found for the innermost ejected particles. The accumulation of particles around $\Delta R_{\rm{final}}=1$ represents cases in which the reconstruction leads to small final radii, i.e., to non-ejection. The accumulation around $10^3$ and above, on the other hand, corresponds to the opposite case where the reconstruction leads to a large final radius for particles that are not ejected in the inline tracking.  For particles with $R_0>3000$~km differences are small and there is a relatively sharp transition from the large differences at small radii to the good agreement on the outside starting from initial radii around 2,500~km. This is in agreement with the increase of the minimum crossing timescales as shown in Figure \ref{fig:timescales}(a). 
Particles with different fate are visible for smaller radii, most noticeably in the region below 2,000~km, where ejected particles are rare. It is interesting to note that none of the particles that are ejected from an initial radius below about 1,500~km are well reproduced, demonstrating that it is impossible to reproduce the exact trajectories of such particles with the limited time resolution of the simulation snapshots.  

Both models, D9.6 and D10.9, show a significant "re-shuffling" between ejecta and non-ejecta particles due to the forward reconstruction, while the total ejecta mass is well reproduced. This illustrates that integrated and average quantities, such as the total mass, can be well reproduced with reconstructed particles, but details of individual trajectories can be radically different. Similar effects are discussed below for the electron fraction in Section~\ref{sec:conditions} and for the integrated nucleosynthesis in Section~\ref{sec:nucleosynthesis}.

\section{Nucleosynthesis conditions}
\label{sec:conditions}
The nucleosynthesis for particles that reach very high temperatures, during which NSE is achieved, is strongly dependent on the electron fraction $Y_e$, which determines the composition when the equilibrium of reactions starts to freeze-out.

For $Y_e\approx 0.5$, which applies to most of the ejecta, the final composition is dominated by \iso{56}{Ni}, as the main product of explosive Si-burning, accompanied by characteristic isotopes such as \iso{44}{Ti} and \iso{40}{Ca} \citep{Woosley.Arnett.ea:1973}.
There is a fundamental difference in the NSE composition, between proton-rich and neutron-rich conditions, as pointed out by \citet{Seitenzahl.Timmes.ea:2008}.
For proton-rich conditions, $Y_e>0.5$, the NSE composition at 5--6~GK remains dominated by Fe, Ni and Cu nuclei as well as free protons and $\alpha$ particles, which 
have only a limited effect on the composition after freeze-out, due to the Coulomb barrier. Only very high $Y_e$, high entropy or strong neutrino exposure allow for significant production of elements heavier than Fe in proton-rich conditions.
For neutron-rich conditions, on the other hand, the dominant nuclei become heavier and the left-over neutrons and $\alpha$ particles allow subsequent nucleosynthesis of heavier elements.
Even for slightly neutron-rich conditions, i.e., $0.46<Y_e<0.50$, the nucleosynthesis proceeds by $\alpha$ captures on neutron-rich Fe-group nuclei up to the closed neutron shell $N=50$, resulting in characteristically high abundances of  \iso{90}{Zr}, accompanied by  \iso{92}{Mo}, \iso{84}{Sr}, \iso{78}{Kr}, \iso{58}{Ni} and \iso{60}{Ni} \citep{Wanajo.Mueller.ea:2018, Hoffman.Woosley.ea:1996,Woosley.Hoffman.ea:1992}. For even lower $Y_e$, more free neutrons are available and heavier nuclei can be produced. 
Due to the key role of $Y_e$ in determining the nucleosynthesis yields, we show in the following how the values of $Y_e$ from reconstructed tracer particles compare to the inline particles, which will be the key for the comparison of the nucleosynthesis results in Section~\ref{sec:nucleosynthesis}. 
Here, we first discuss model D9.6 in detail and show that D10.9 follows similar trends, but with much better sampling due to the larger number of tracer particles in the ejecta.
\begin{figure}
    \centering
    \includegraphics[width=\linewidth]{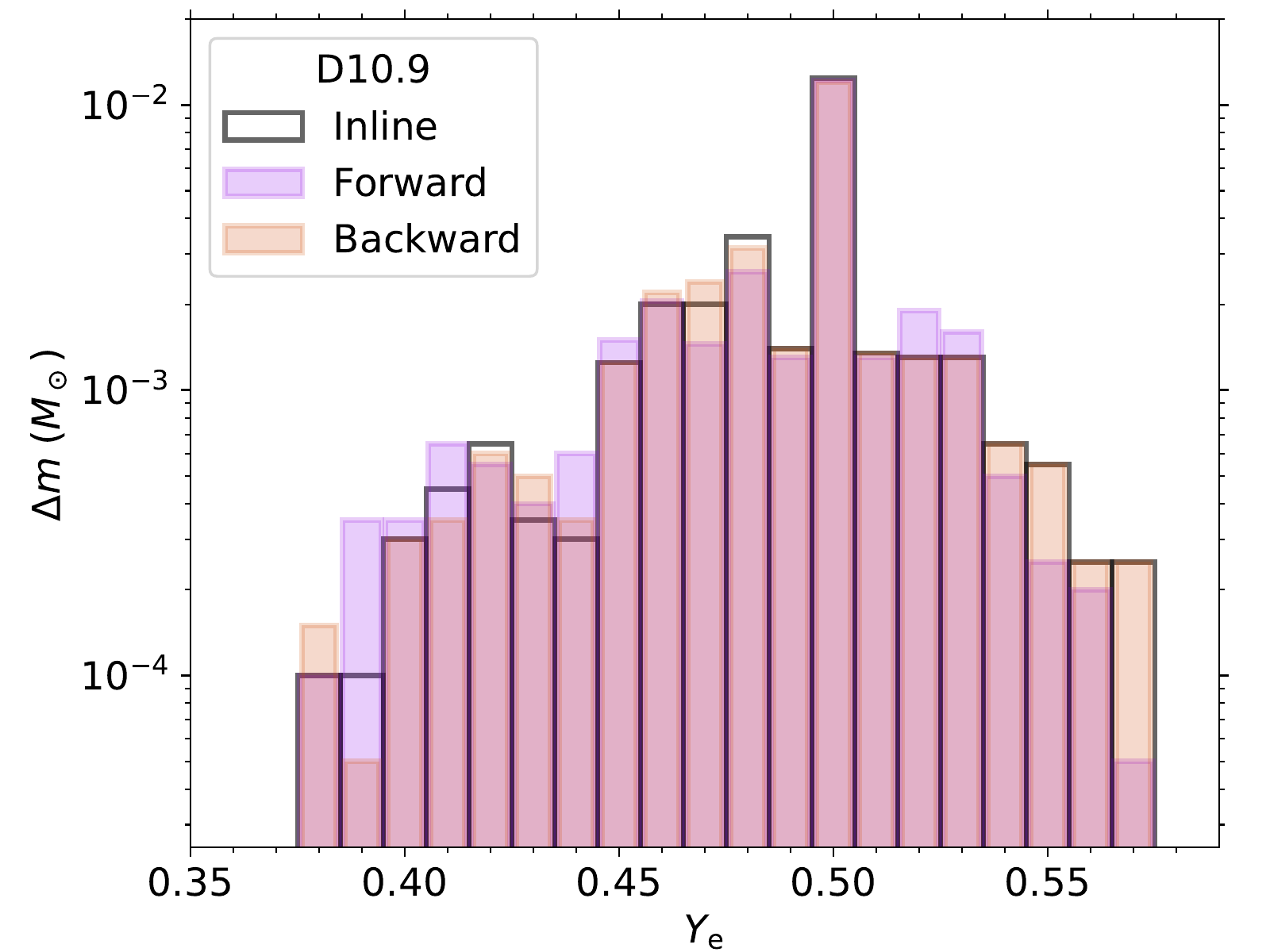}
    \caption{Distribution $Y_e$ at freeze-out (8GK) for the ejected tracer particles from model D9.6. The distribution of the inline particles is shown as black, empty boxes, while the excesses of the forward and backward reconstructed particles are shown in light blue and light orange, respectively. The overlap between the two reconstruction methods is shown in pink. The mass per particle is $5\times 10^{-5}\,M_{\odot}$. }
    \label{fig:ye_hist_D96}
\end{figure}

Figure \ref{fig:ye_hist_D96} shows the $Y_e$ distribution from model D9.6 for the ejected tracer particles at the time when the temperature drops below 8~GK, or at the time the peak temperature is reached for those particles that never reach NSE. The overall shapes of the distributions from the inline and reconstructed particles are very similar, but there are clear differences in the tails of the distributions, that are not well sampled.
At low $Y_e$, there is a noticeable excess of particles from the forward reconstructed set, most prominent in the bin at $Y_e=0.39$. 
At the upper end of the distribution, on the other hand, the number of tracers
from the forward integration is noticeably below the number of inline tracers. 
In contrast, the number of particles in the high $Y_e$ tail from the backward reconstruction noticeably exceeds the number of inline tracers.
Overall, the distribution from the backward reconstruction is visibly closer to the distribution of the inline particles at both extremes of the distribution.

\begin{figure}
    \centering
    \includegraphics[width=\linewidth]{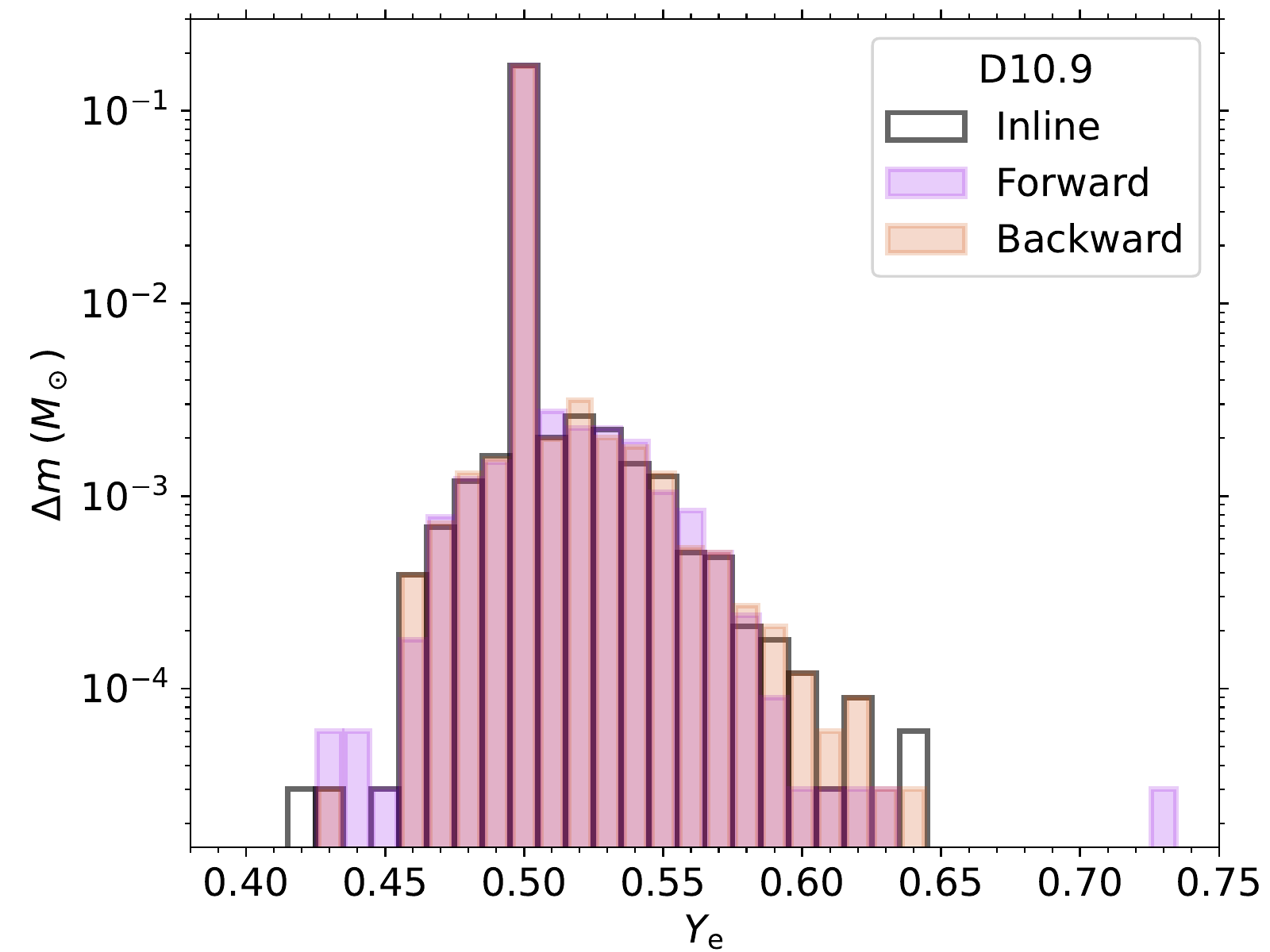}
    \caption{Same as Figure \ref{fig:ye_hist_D96}, showing the distribution of $Y_e$ of the ejected particles from model D10.9. Note that the range of $Y_e$ is different from Figure \ref{fig:ye_hist_D96} but the size of the $Y_e$ bins is the same. The ejecta mass of model D10.9 is also significantly larger than for model D9.6. The mass per particle is $3\times 10^{-5}\,M_{\odot}$.}
    \label{fig:ye_histo_d109}
\end{figure}

\begin{figure}
    \centering
    \includegraphics[width=\linewidth]{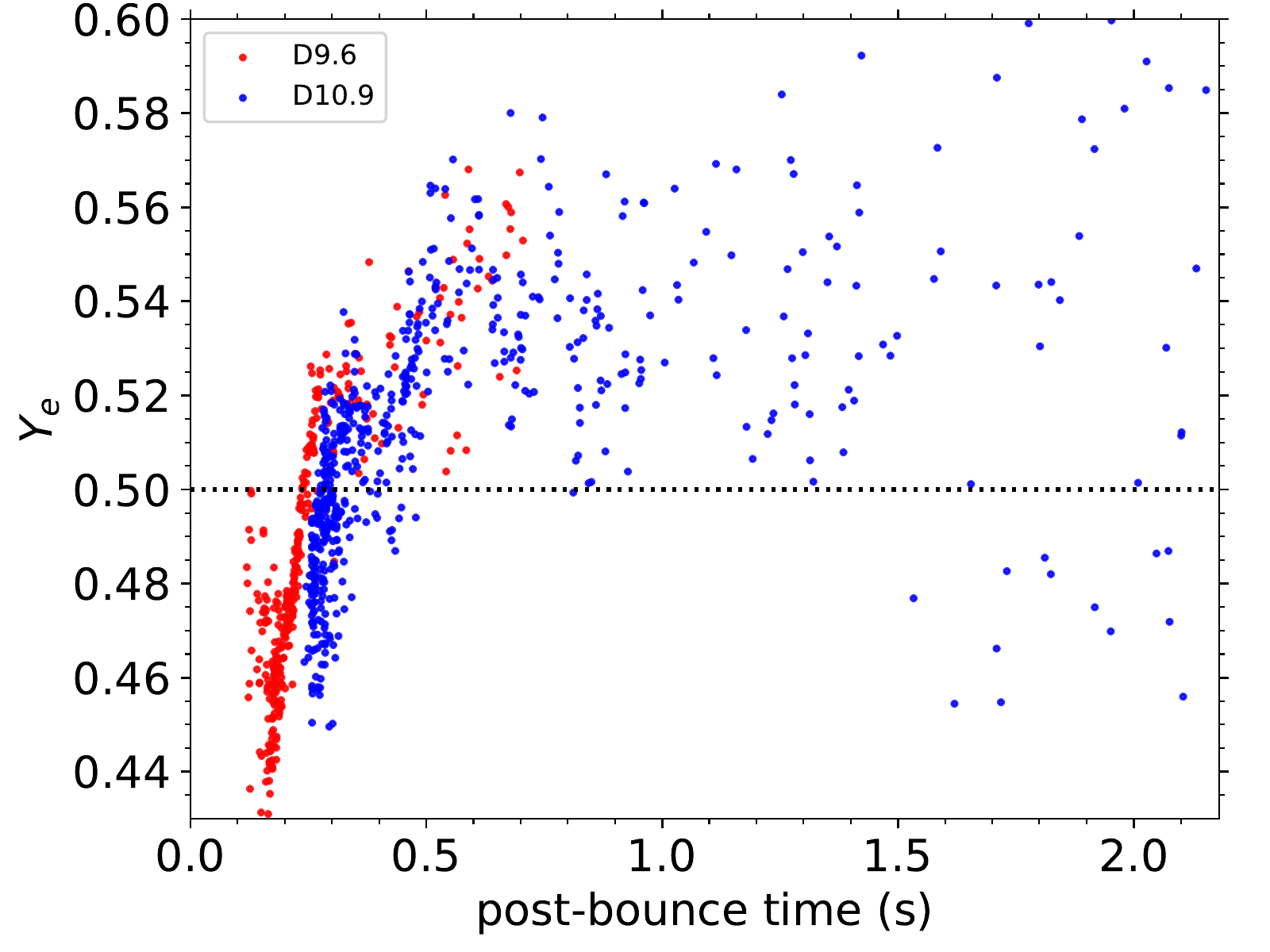}
    \caption{Electron fraction at freeze out (8 GK) as a function of the post-bounce time of freeze-out for each particle for models D9.6 and D10.9. There is a clear increase of $Y_e$ with time in the early phase, with some large scatter at late times in D10.9. }
    \label{fig:ye_vs_time}
\end{figure}

Figure \ref{fig:ye_histo_d109} shows the histogram for the $Y_e$ distribution of the tracer particles for model D10.9, comparing the forward and backward reconstruction to the inline tracked particles.  
The bulk of the ejecta between $0.47<Y_e<0.58$,  sampled by many tracer particles, seems well reproduced. The tails of the distribution, however, again show that the forward reconstruction tends to underestimate the amount of the most proton-rich ejecta while the backward reconstruction misses the few tracers that experience 
the lowest $Y_e$. This is similar to the trends for model D9.6. The only exception is an extremely proton-rich particle that freezes out with $Y_e=0.73$ in the forward reconstruction. This particle samples an unusual late-time, low-mass proton-rich outflow that is sampled by neither the inline particles nor the backward reconstruction. 
Except for this chance particle, the systematic trend of the forward reconstruction to underestimate the proton-rich tail of the distribution, can be explained by the timing of the different ejecta components illustrated by the correlation between the time of freeze-out and the value of $Y_e$.
The proton-rich ejecta are usually late-time ejecta that have been subject to significant neutrino processing, while the most neutron-rich material tends to be ejected early, before significant neutrino irradiation takes place. This trend is shown in Figure \ref{fig:ye_vs_time} for the inline tracked particles of both models, although there is  large scatter at late times in D10.9.
On average, the time of freeze-out of proton-rich material is relatively close to the end of the simulation and the backward reconstruction only needs to accurately reproduce a relatively short fraction of the trajectory. The forward reconstruction, on the other hand, would need to follow the particles accurately through one or multiple phases of strong neutrino heating, in many cases close to the PNS where timescales are short. This makes integration errors much more likely. Part of the highest-$Y_e$ ejecta also tends to be classified as non-ejecta in the forward reconstruction as we will show below. The lowest $Y_e$ ejecta, on the  other hand, are typically ejected early, without much neutrino processing. Here, the backward reconstruction is at a disadvantage, because it would need to trace the trajectory accurately over a longer time period. The trajectories of the early ejecta, however, do not include convective overturn or complicated motion induced by neutrino heating, which makes it easier for the backward integration to also capture the neutron-rich tail of the distribution, giving an overall better match compared to the inline tracked particles.

\begin{figure}
    \centering
    \includegraphics[width=\linewidth]{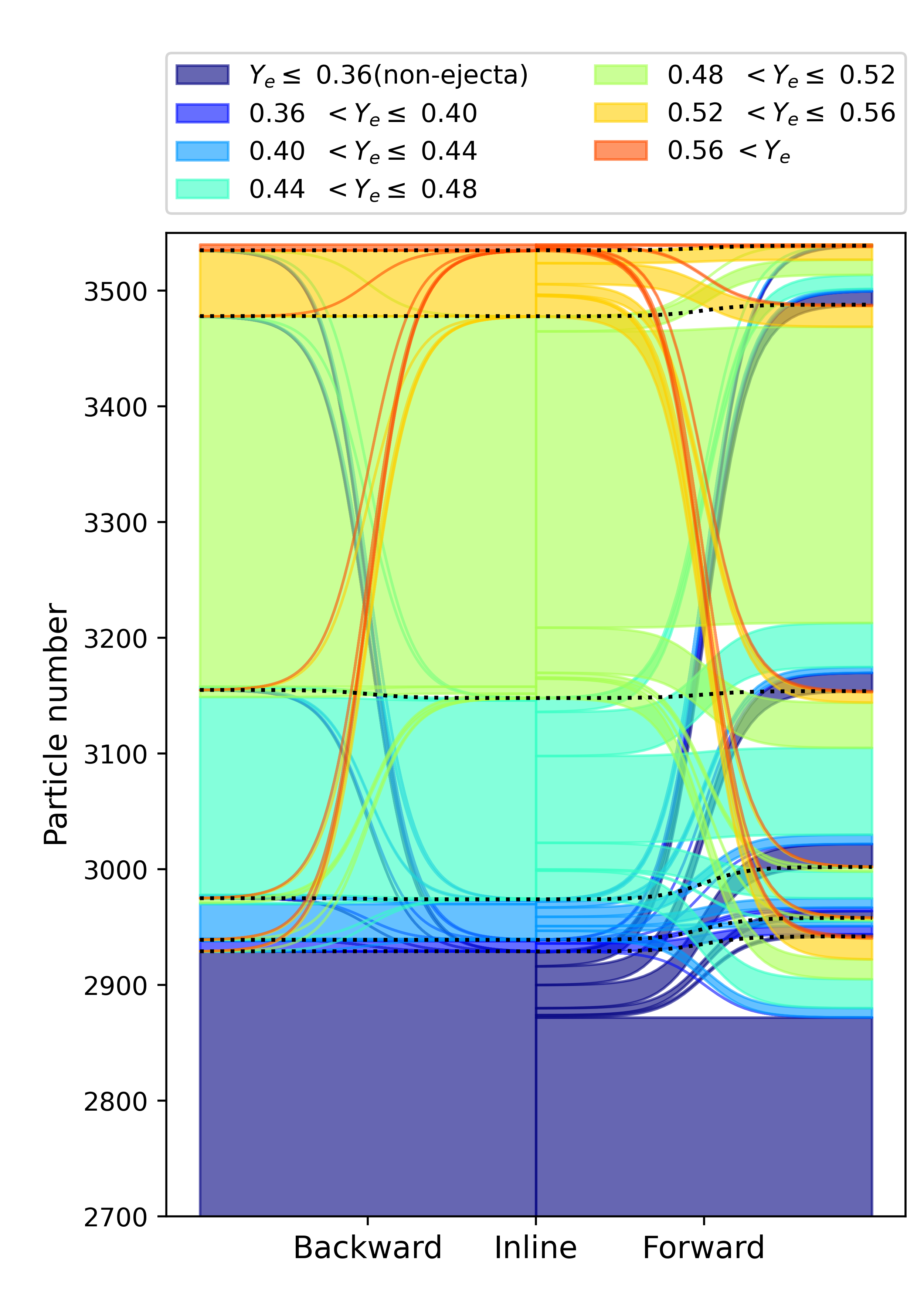}
    \caption{Movement of particles between bins of $Y_e$ at freeze-out (8 GK) when using reconstruction instead of inline particles from model D9.6. From the center, the bands indicate which $Y_e$ bin particles belong to when the forward (right hand side) or backward (left hand side) reconstruction is used. The width of the bins indicates the particle number that is shown cumulatively on the ordinate axis. Particles that are not ejected are all in the lowest $Y_e$ bin. Note that the positions of the backward reconstructed particles at the end of the simulation are identical to the inline particles by construction and, thus, there is no change in whether a particle is ejected or not. In contrast, the number of ejected particles changes when the forward reconstruction is used.}
    \label{fig:ye_migration}
\end{figure}

In the following, we look more closely at the differences for model D9.6.
There is a clear association between particles from the inline tracking and the two reconstruction approaches and we can connect the $Y_e$ of the inline particle to the corresponding reconstructed particle to make comparisons on a particle-by-particle basis.
Figure~\ref{fig:ye_migration} illustrates how individual particles move between $Y_e$ bins when comparing the reconstruction methods for model D9.6. The shaded regions in the center indicate the distribution of the inline tracer particles in $Y_e$ at freeze-out with color-coded bins. The colors are assigned with respect to the inline tracked particles. The dotted horizontal lines indicate the separation between $Y_e$ bins, with the low $Y_e$ at the bottom and high $Y_e$ at the top, as defined by the colors in the center. The y axis indicates the particle count. Particles that are not ejected are grouped with the lowest $Y_e$ bin. Thus, only the 603 ejected particles are in the upper bins. 
Towards the left and the right, the colors from the original $Y_e$ remain, but the position corresponds to the $Y_e$ bin of the same particle in the reconstruction. The color composition of the bins on the left and the right thus represents the "origin" of the tracers in terms of the $Y_e$ in the inline set as they fill up the bins in the reconstructed sets. The shaded bands thus indicate how individual particles move between $Y_e$ bins depending on the direction of the reconstruction. In this way, Figure~\ref{fig:ye_migration} shows that there is relatively  little motion of particles between bins towards the backward integration on the left. Towards the right, however, there is some significant movement between bins for the forward reconstruction.
Furthermore, while the number of ejected particles is identical for the backward reconstruction by definition, there is a noticeable exchange between particles that are ejected in the inline tracer and not in the forward reconstruction and vice versa. This re-shuffling is also discussed in Section~\ref{sec:final_fate} and it is more clearly visible here. The total number of ejected particles, however, changes only slightly.
It is interesting to note that the motion is not limited to adjacent bins, but there are significant exchanges across the whole range of $Y_e$ which indicates drastic changes of individual trajectories. This applies, in particular, to the particles that have originally not been ejected in the simulation and appear in every $Y_e$ bin in the forward reconstruction, distributed roughly proportional to the number of particles in the bin for the inline particles. 
Figure~\ref{fig:ye_migration} also reveals that the largest movement away from the highest $Y_e$ bin goes to the non-ejecta. This shows that the particles that are missing in the ejecta in the forward reconstruction in the high $Y_e$ tail of Figure~\ref{fig:ye_hist_D96} are not ejected in the reconstruction. The high $Y_e$ indicates that these particles require significant neutrino irradiation and thus heating, which suggests that they are part of the late-time, wind-like ejecta. We can furthermore see that the slight excess in low $Y_e$ particles results from an influx of particles from the higher $Y_e$ and from part of the non-ejecta, representing particles that originally were caught in convective overturns, but that get ejected early on in the reconstruction by missing the downward turn.

\begin{figure}
    \centering
    \includegraphics[width=\linewidth]{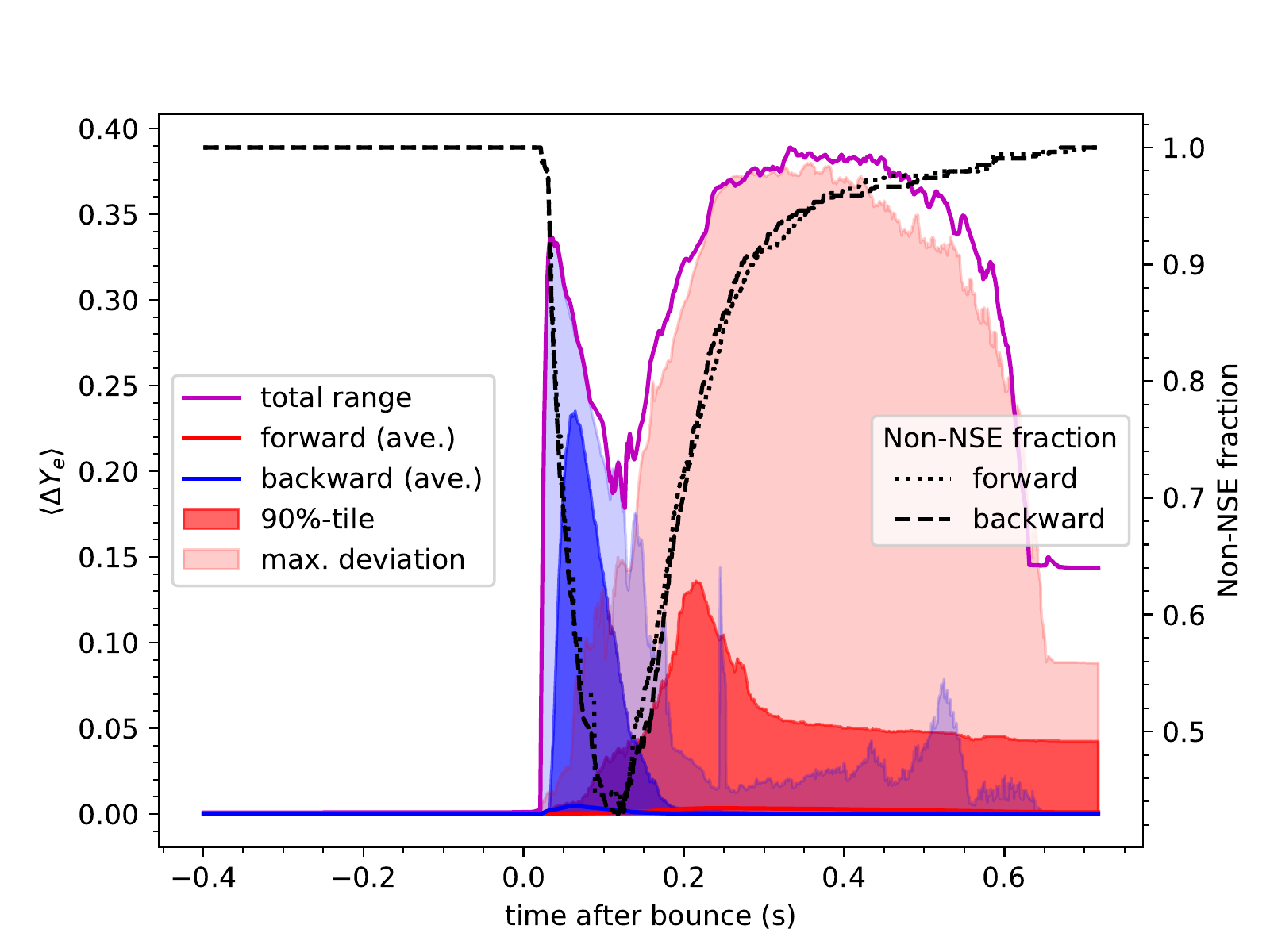}
    \caption{Quantiles of particle-by-particle deviation $\Delta Y_e$ from the inline tracers from D9.6. The light shaded region shows the full range of values of the deviation, compared to the range of values of $Y_e$ for the inline tracers as magenta line. The dark shaded region indicates the 90th-percentile of the distribution, i.e., \percent{90} of the tracer particles show deviation that are smaller than the indicated value. }
    \label{fig:ye_dev}
\end{figure}

To quantify how the backward reconstruction gives a more reliable estimate of the $Y_e$ at freeze-out, we define
 for particle $i$
 \begin{equation} \label{eq:delta_ye}
 \Delta Y_{e}^{i}=|Y_{e}^{i,\rm{reconstr.}}-Y_{e}^{i,\rm{inline}}| 
 \end{equation}
 with the the value $Y_{e}^{i,\rm{reconstr.}}$ from the forward or backward reconstructed tracer particles and $Y_{e}^{i,\rm{inline}}$ from the inline particles. We obtain a value of $\Delta Y_e^i$ for each particle and key quantities of the distribution are shown as a function of time in Figure \ref{fig:ye_dev}, taking only particles that become part of the ejecta in all cases. For reference, we also show the whole range of $Y_e$ that is covered by all the inline tracer particles and varies significantly with time. The black dashed and dotted lines indicate the fraction of tracer particles that are not in NSE, i.e., have a peak temperature below $8\,\rm{GK}$. 
Before bounce, there are no noticeable differences because the range of values of $Y_e$ at the initial positions of the ejected 
tracer particles is very narrow, ranging from $0.4992$ to $0.5$, corresponding to the $Y_e$ profile of the pre-SN model in the region where the tracers are originally placed. Initially, almost all of the ejected tracer particles are from non-NSE regions. 
The range of available $Y_e$ and also the deviations start to increase around 20 ms after bounce. This corresponds to the time when the first tracer particles reach NSE and are exposed to neutrinos. 
Neutrinos are much more effective at changing the value of $Y_e$ when the composition is dominated by free protons and neutrons. 
Once shock and neutrino heating induce high temperatures and dissociation into free nucleons, the available range of $Y_e$ rapidly expands. 
Therefore, Figure \ref{fig:ye_dev} shows that the range of available $Y_e$ expands when the non-NSE fractions reaches a minimum. 
The maximum differences of the forward reconstruction fills out almost the whole available range of $Y_e$, which is largely the result of the re-shuffling between ejecta and non-ejecta. For example, some particles that are just shock-heated remain at $Y_e\approx 0.5$ in the inline tracking, fall all the way to the PNS in the reconstruction, giving rise to large deviations in $Y_e$ for individual tracers. At the end of the simulation, however, the maximum difference is around 0.08, which is half of the available range, indicating that many particles reach similar conditions eventually, even though the detailed trajectory varies drastically. The 90th percentile also shows that there are still \percent{10} of the particles that differ by more than 0.05 at the end of the simulation with the forward reconstruction. 
With the backward reconstruction, the trends are in the opposite direction. The differences increase sharply at times before $\approx 200$~ms after bounce and even fill out the whole range of available $Y_e$ around bounce. On the other hand, \percent{90} of the particles show only very small difference for later times. Even the maximum differences do not fill out the available range and are thus much smaller compared to the forward reconstruction. One key aspect here is that the differences are also small when the non-NSE fraction increases, i.e., when most of the particles freeze-out, which is crucial for the nucleosynthesis outcome.

Based on the two models that we have considered here, we can conclude that there is a systematic trend for forward reconstruction to slightly underestimate the amount of highest $Y_e$ ejecta. The opposite is true for the backward reconstruction due to the general trend of high $Y_e$ ejecta to emerge at later times than the early, low $Y_e$ ejecta in CCSNe.

\section{Nucleosynthesis yields}
\label{sec:nucleosynthesis}
In the following we present and compare the integrated nucleosynthesis yields, first for model D9.6 and then for model D10.9. To judge the relevance of the production of an isotope, it is useful to compare the mass fraction to the solar value of the same isotope. This ratio, $P(A,Z)=X(A,Z)/X_\odot(A,Z)$, is referred to as the production factor. Note that, in the context of chemical evolution, an astrophysical site can only be expected to significantly contribute to
the solar inventory of a given isotope if its production factor is close to the maximum of the production factors for this site.
 In this article, we will not discuss any implication of the nucleosynthesis yields for observations or chemical evolution in detail, but focus on the comparison between the different approaches of extracting tracer particles. A dedicated study of the nucleosynthesis of all D-Series models will be published separately.
The simulation domain for both models covers the the stars out to the lower He~shell. Additional ejecta from the He~shell outward and parts of the still-to-be shocked C~shell will still enhance the production factors of the lighter elements.

\subsection{Model D9.6}
\begin{figure*}
\centering
\includegraphics[width=\textwidth]{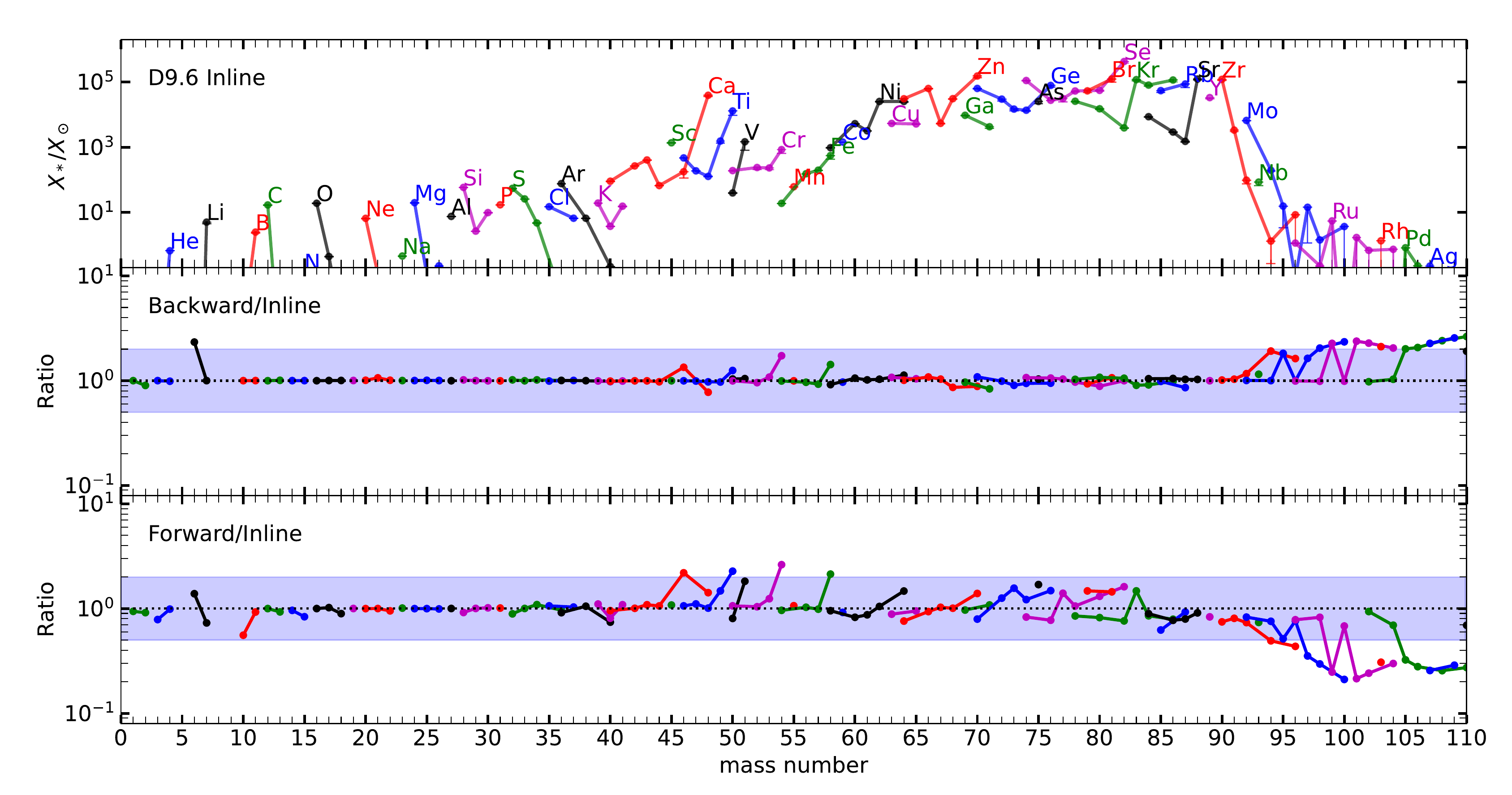}
\caption{Production factors from the inline tracer particles for model D9.6 are shown in the top panel. The downward error bars indicate the magnitude of the contribution of the single tracer particle with the largest mass fraction of the corresponding isotope. The production factor for the backward (middle panel) and forward (bottom panel) reconstructed particles relative to the inline data. The blue shaded bands indicate the range of a factor two around a ratio of one.}
\label{fig:pf_comparison}
\end{figure*}
The most important products from CCSNe from the point of view of observation are the radioactive isotopes \iso{56}{Ni} and \iso{44}{Ti}. With the inline particles, we find $5.18\,\times 10^{-3}\msun$ of \iso{56}{Ni} to be produced at the end of the nucleosynthesis calculations which extrapolate the trajectories to 100~s after core bounce. This is well reproduced by the reconstructed particles 
with a relative difference with respect to the inline particles of \percent{3.7} for the backward reconstruction and \percent{2.2} with the forward reconstruction. The agreement for \iso{44}{Ti} is similarly good, with a relative difference of \percent{2} and \percent{6} for the backward and forward reconstruction, respectively. The yield of \iso{44}{Ti} is rather low for this low mass model with $2.78\,\times 10^{-6}\msun$, based on the inline particles. 

For the stable isotopes, the top panel of Figure \ref{fig:pf_comparison} shows the production factors for the nucleosynthesis yields of model D9.6 based on the inline tracer particles. Here we have followed decay chains for all radioactive isotopes to the final, stable product.
Overall, the nucleosynthesis pattern agrees well with previous evaluations by \citet{Wanajo.Mueller.ea:2018} of the same progenitor.
Due to its steep density gradient, this model explodes rapidly leading to a significant amount of early, neutron-rich ejecta, with significant yields of \iso{48}{Ca} and the neutron-rich isotopes of Ni, Zn, Ge, Se and Kr. \iso{90}{Zr} and the $p$-nuclei \iso{78}{Kr}, \iso{74}{Se} and \iso{84}{Sr}, which have very low solar abundances, also exhibit significant production factors. 
The later, proton-rich ejecta contribute mostly additional  Fe-group elements, including \iso{50}{Cr} and \iso{54}{Fe}, as well as isotopes such as \iso{45}{Sc} and \iso{39}{K}.
Comparing our results to Figures 10 and 11 of \citet{Wanajo.Mueller.ea:2018}, we find very similar results for the correlations between final compositions as function of freeze-out $Y_e$. 

The bottom two panels of Figure~\ref{fig:pf_comparison} show the ratios of the results with the reconstructed tracer particles to the inline particles for model D9.6. 
Overall, the ratios show that the backward reconstruction is in better agreement with the inline tracer particles, which is as expected from the discussion of the nucleosynthesis conditions in Section~\ref{sec:conditions}. 
In the following, we look at the largest deviations and give more detailed explanations for the origin of the differences, demonstrating the sensitivity of isotopes that are produced predominantly by small subsets of particles.

\paragraph{\iso{50}{Ti}, \iso{54}{Cr} and \iso{58}{Fe}}
Backward and forward reconstruction produce an excess of \iso{50}{Ti}, \iso{54}{Cr} and \iso{58}{Fe} compared to the inline tracer particles. Among the isotopes with mass number $A<91$, these are among the largest discrepancies. The three isotopes have $Z/A\approx 0.44$ and are consequently produced from NSE in a narrow range of $Y_e$ from 0.43 to 0.45. Figure \ref{fig:ye_hist_D96} shows that both types of reconstruction lead to slightly more material freezing out under these conditions as discussed in Section~\ref{sec:conditions}. 
More material freezing out with $Y_e$ in this range directly leads to the excess of these isotopes.
Figure \ref{fig:ye_hist_D96} also shows that these conditions are represented by only 4-5 particles per $Y_e$ bin for the inline tracers, indicating that isotopes that are produced by a small number of particles are the most sensitive. 

With the inline tracers, half of the production of \iso{54}{Cr} results from 4 particles and the conditions of these 4 particles also dominate the yields of \iso{50}{Ti} and \iso{58}{Fe}.  
These 4 particles are well reproduced by the backward reconstruction. 
However, the backward reconstruction results in 2 additional particles in the optimal $Y_e$ range that are then responsible for more than half of the yield, reaching \iso{54}{Cr} mass fractions of $0.079$ and $0.036$ due to values of $Y_e$, $0.43$ and $0.42$, at freeze-out. 
The corresponding two particles in the inline set, do not produce any \iso{54}{Cr} because they freeze out from NSE with values of $Y_e$ close to $0.5$.

\begin{figure}
\centering
\includegraphics[width=\linewidth]{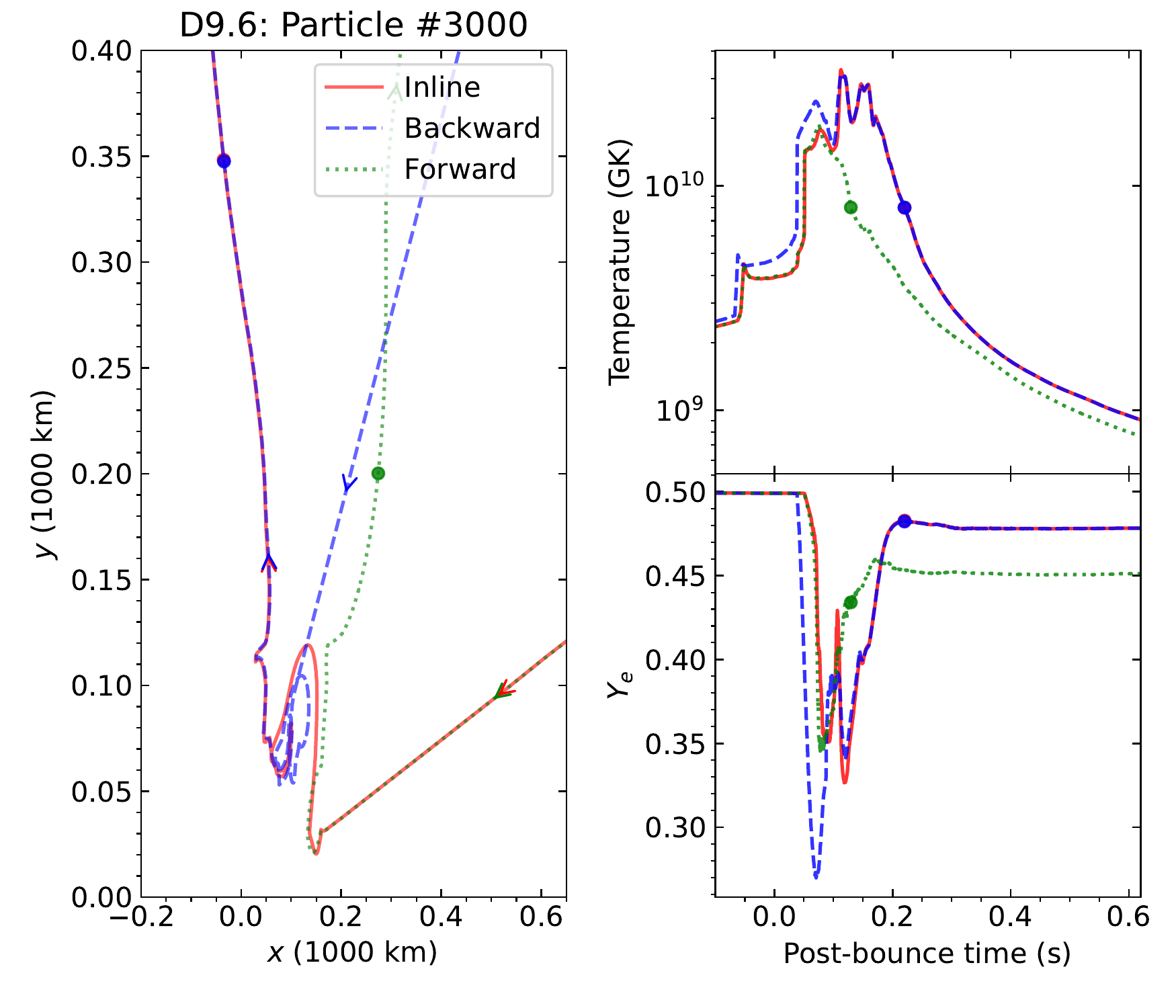}
\caption{\label{fig:p3} Spatial trajectories as well as temperature and $Y_e$ evolution for a particle that dominates the production of \iso{54}{Cr} with the forward reconstruction. The evolution is clearly different for both types of reconstruction, but the freeze-out $Y_e$ changes most with the forward reconstruction.} 
\end{figure}

In forward tracing, one of the particles that makes a major contribution to \iso{54}{Cr} in the inline case is not ejected at all and the three other tracers that dominate the production for the inline case show significantly reduced mass fractions. 
Instead, half of the \iso{54}{Cr} from the particles reconstructed with forward integration, is represented by two other particles.
The trajectory of one of these particles is shown in Figure \ref{fig:p3}. 
Close inspection of the trajectory in space reveals that the inline tracked particle undergoes several convective turnovers in the region between 50 and 100 km. 
During this phase, both the forward and backward reconstruction deviate from the inline particle. 
For the backward reconstructed particle, this leads to a very different initial position, but the freeze-out conditions are still very similar. 
In contrast, the forward reconstructed particle does not get trapped in the convective motion, but gets ejected straight away, leading to an earlier freeze-out at a lower value of $Y_e$. 
 
 \paragraph{Neutron-rich Ca}
 \iso{46}{Ca} is produced by the four most neutron-rich particles with $Y_e<0.4$, making the yield very sensitive to small changes.
 The backward and forward integrated particles result in \percent{35} and \percent{100} increases in the yield, respectively. 
 For the backward integrated particles, this is mostly due to the reduction of the freeze-out value of $Y_e$ for one particle from 0.40 for the inline particle to 0.38 for the reconstructed particle due to a spatial displacement of around 150 km. This makes it the particle with the largest \iso{46}{Ca} mass fraction, $7 \times 10^{-5}$, in all three particle sets for this model. 
 Among the forward-reconstructed particles, the largest contribution is from one particle that does not produce a significant amount of \iso{46}{Ca} in the backward integrated and inline sets, where it remains close to the PNS until 300 ms after bounce, subject to strong neutrino irradiation. Consequently, it freezes out from NSE with $Y_e=0.48$. The forward integration sets the particle on a trajectory that already freezes out from NSE at 100 ms after bounce with $Y_e=0.38$, leading to a high mass fraction of $5 \times 10^{-5}$ for \iso{46}{Ca}.
 
The yield of \iso{48}{Ca} corresponds to one of the largest production factors for this model and a high yield of  \iso{48}{Ca} was also found by \citet{Wanajo.Mueller.ea:2018} for the same progenitor model.  
Despite the large production factor, the yield is not very well resolved, with only 8 particles responsible for more than half of the total yield. 
 The production is dominated by particles that freeze-out from NSE with $Y_e$ in a narrow range around 0.4. Due to this narrow range, its yield is underestimated by the backward reconstruction by about \percent{20} compared to the inline-tracked particles and overestimated by \percent{50} with the forward reconstruction.

 \paragraph{Zn to Zr}
 Elements from Zn to Zr dominate the production factors for model D9.6. Due to the presence of both neutron-rich and proton-rich ejecta, all isotopes of these elements are produced at a similar level. This is very similar to the results reported by \citet{Wanajo.Mueller.ea:2018}, who also studied the nucleosynthesis of this stellar model based on axisymmetric simulations.
 
The nucleosynthesis yields of these elements based on the backward integrated particles agree with the inline particles on the order of \percent{10}--\percent{20}, as shown in the middle panel of Figure \ref{fig:pf_comparison}. The bottom panel of the same figure shows that the results using the forward integrated particles show larger differences, albeit not exceeding a factor of two.
The trends in the isotopic chains directly reflect the the trends in the $Y_e$ distribution.
For the isotopic chains Zn, Ge and Se, the most neutron-deficient isotopes in each chain are produced less in the forward reconstruction compared to the inline tracer particles and the most neutron-rich isotopes are produced more. This directly reflects the slight deficiency in proton-rich material with $Y_e\geq 0.53$ and excess in material with $0.38<Y_e<0.49$ shown in Figure~\ref{fig:ye_hist_D96}. With the backward reconstruction, the most neutron-rich isotopes are produced less, reflecting the deficiency of neutron-rich particles.
\iso{90}{Zr} at the closed neutron shell $N=50$ is a nucleosynthesis bottleneck of the $\alpha$ process for values of $Y_e$ up to 0.49 \citep{Hoffman.Woosley.ea:1996}. Therefore, its production is quite robust with respect to small variations of the conditions and both, forward and backward reconstructed particles agree with the inline particles to within \percent{20}.  

\begin{figure}
    \centering
    \includegraphics[width=\linewidth]{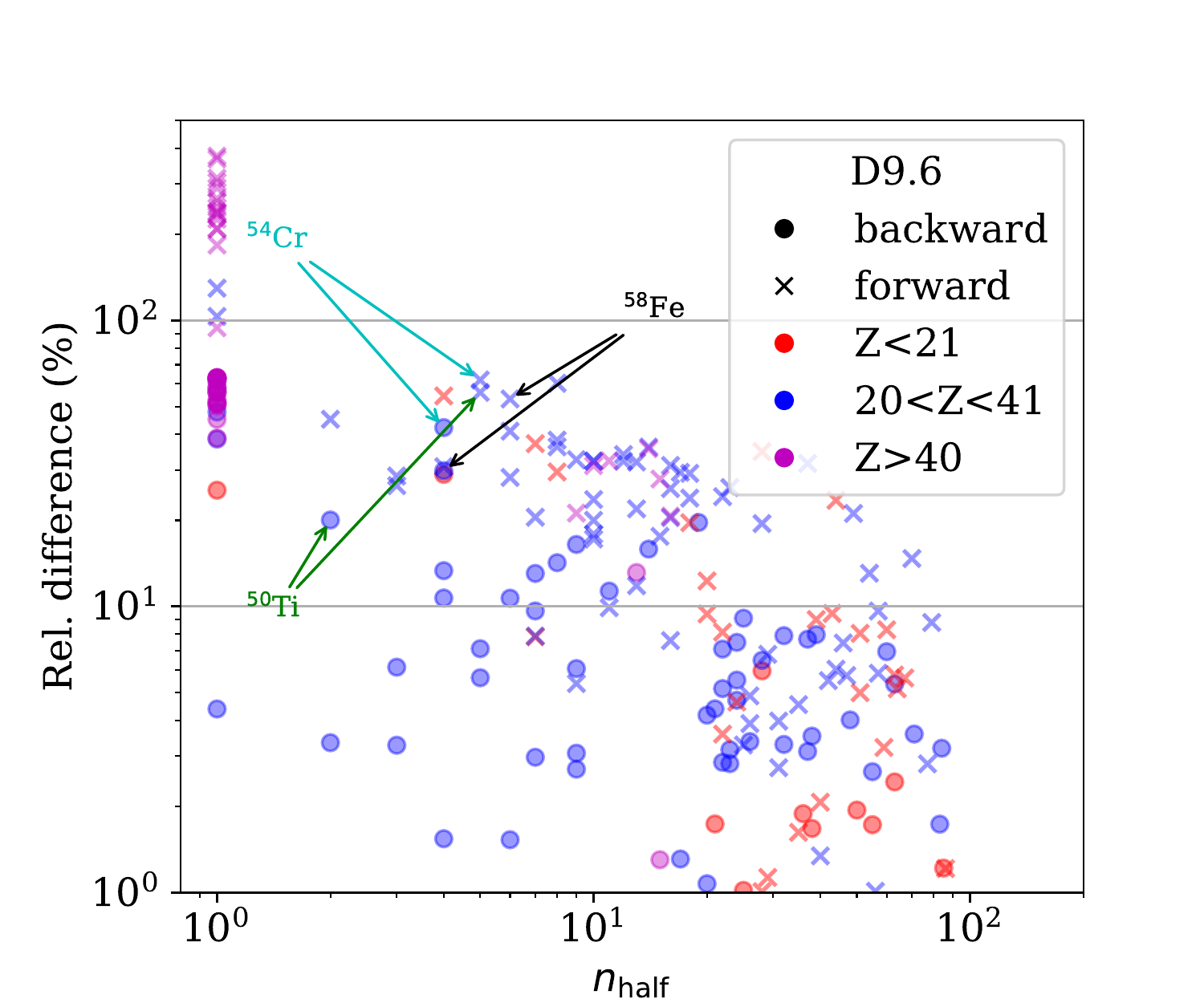}
    \caption{Correlation between relative deviation from the results with the inline tracers and the number of particles that represent half of the yield for each isotope. Deviations are larger for isotopes that are represented by few particles, which are mostly heavier elements and the most neutron-rich isotopes of the Fe-group. }
    \label{fig:nhalf_D96}
\end{figure}

While it is not surprising, the discussion above illustrates that the yields of isotopes that are produced only by a few exceptional particles are the most sensitive to the reconstruction method. 
It also highlights that, because of the wide range of the solar abundances, it is possible, even likely, to have isotopes sourced from only a few tracers still have large production factors and therefore potentially contribute significantly to Galactic chemical evolution.
To quantify how many particles are necessary to consider the yield as robust, Figure~\ref{fig:nhalf_D96} shows the relative difference of the yields between the reconstructed and the inline particles correlated to the number of particles that represent at least half of the yield of this isotope, $n_{\rm{half}}$. 
This visualizes that the largest deviations can be found for isotopes for which only a few particles dominate the total yields. 
The isotopes \iso{50}{Ti}, \iso{54}{Cr} and \iso{58}{Fe} that are discussed above are highlighted.
Elements beyond Zr, i.e., with $Z>40$ are also strongly affected. The yields of these isotopes are dominated by the tails of the $Y_e$ distribution sampled by a few particles and thus deviate by a factor of two or more.
Figure~\ref{fig:nhalf_D96} also shows that for any isotope with $n_{\rm{half}}>20$, the relative difference is below \percent{10} for the backward reconstructed particles. For the forward reconstruction, there are differences larger than \percent{10} also for isotopes with  $n_{\rm{half}}>20$, but differences remain limited to \percent{30}.
\begin{figure*}
    \centering
    \includegraphics[width=\linewidth]{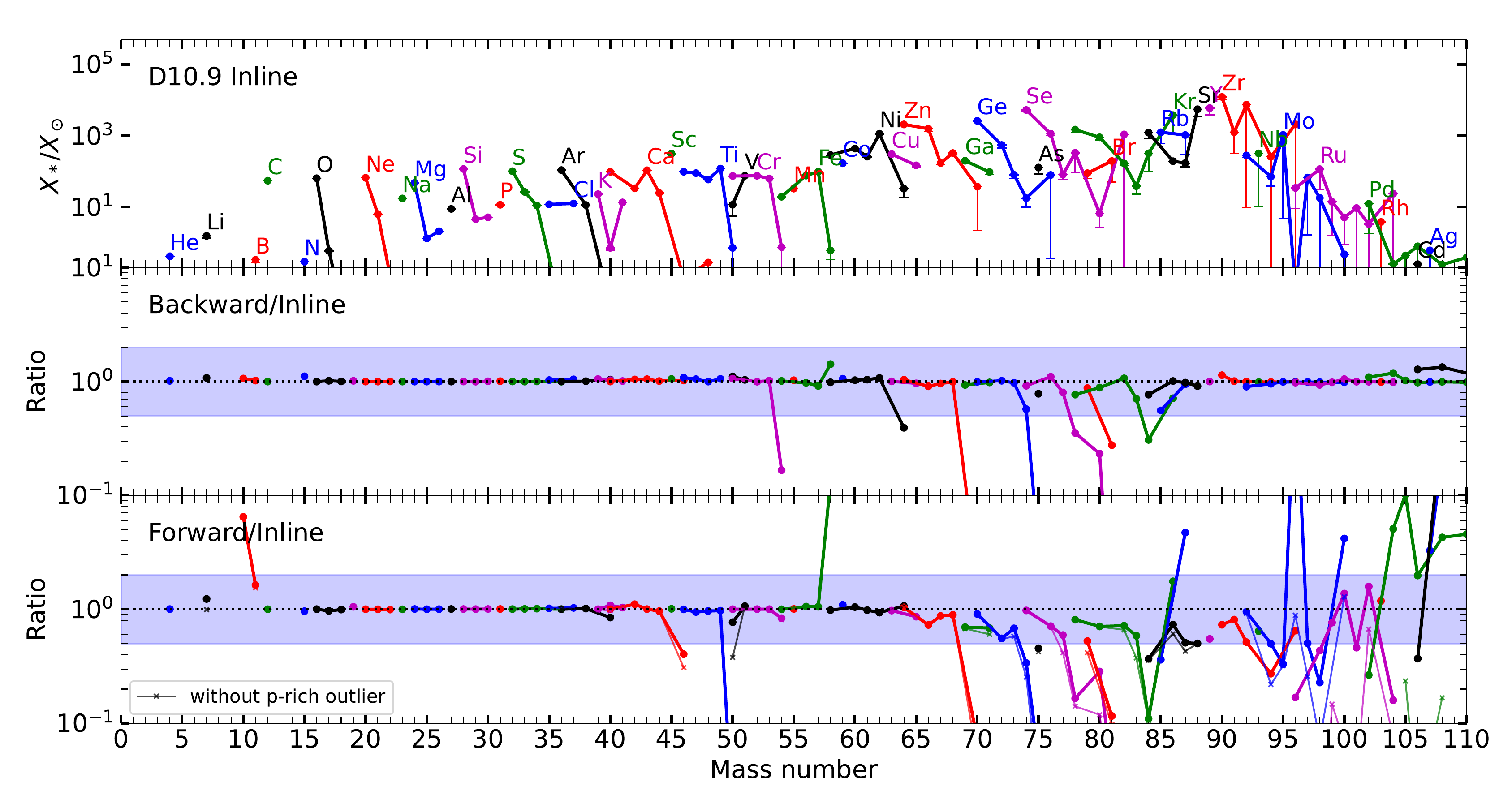}
    \caption{Same as Figure \ref{fig:pf_comparison} for model D10.9-HW10. The thin lines marked with "x" in the bottom panel also show the ratio of the production factor when excluding the particle with the largest value of $Y_e$ from the total yield. }
    \label{fig:pf_comparison_d109}
\end{figure*}
For nucleosynthesis calculations based only on reconstructed particles, it is thus recommended to consider a potentially large uncertainty for yield predictions of isotopes for which half the yield results from less than 20 particles. 

\subsection{Model D10.9}
With the higher explosion energy and ejecta mass, model D10.9 yields $1.37\times 10^{-2}$~\msun\ of \iso{56}{Ni} based on the inline tracers. The relative difference to this for the backward and forward reconstructed particles is \percent{2} and \percent{6}, respectively. The yield of \iso{44}{Ti} is low, $5.51\times 10^{-6}$~\msun\ with the inline particles and the relative difference with reconstructed particles are \percent{1} and \percent{4} for backward and forward integration, respectively.

The top panel of Figure \ref{fig:pf_comparison_d109} shows the production factors for model D10.9 based on the inline tracer particle data.
In contrast to D9.6, model D10.9 ejects less neutron-rich material which is represented by only a small number of particles. These particles still have a very strong impact on the yields, leading to large production factors for \iso{90}{Zr} and \iso{88}{Sr}, which require neutron-rich conditions.
Due to the proton-rich component, however, there is also a clear enhancement of the most neutron deficient isotope relative to the most neutron-rich ones along each isotopic chain and the production factors are less flat than for model D9.6. The proton-rich conditions allow for relatively large production 
factors for the $p$-nuclei
\iso{74}{Se}, \iso{78}{Kr} and \iso{84}{Sr}. Furthermore, there is a characteristically large production of \iso{45}{Sc} \citep{FrHaLi06}. This pattern is similar to
yields predicted from supernova simulations in this mass range \citep{Wanajo.Mueller.ea:2018,Sieverding.Mueller.ea:2020,Witt.Psaltis.ea:2021}

Overall, the yields based on the backward integrated particles show a very good agreement with the original tracer particles as shown by the ratios in the middle panel of Figure \ref{fig:pf_comparison_d109}. With a few exceptions, the yields of all isotopes match to within significantly less than a factor two.
Similar to D9.6, neutron-rich isotopes with very low production factors, such as \iso{54}{Cr}, \iso{58}{Fe}, \iso{63}{Ni}, \iso{75}{Ge}, \iso{78,80}{Se}, \iso{81}{Br} and \iso{84}{Br} deviate by more than a factor two. All of these isotopes are on the neutron-rich side of their isotopic chains and require $Y_e<0.45$ to be produced. Their yields are, thus, very low for this model with predominantly proton-rich ejecta and the yields are dominated by very few particles. The large differences in the integrated yield can be explained by the differences in the distribution of $Y_e$ at freeze-out discussed in Section~\ref{sec:conditions}. For example, \iso{84}{Kr}, which is predominantly produced as \iso{84}{Br} in neutron-rich conditions with $Y_e\approx 0.41$ is dominated by a single particle that freezes out with the right $Y_e$ to produce this isotope in the inline tracked particles. This particle is not well reproduced in the backward reconstruction, where it turns out to be proton-rich, freezing out with $Y_e=0.52$. This missing particle alone reduces the \iso{84}{Kr} yield by more than a factor three. 
Figure \ref{fig:ye_histo_d109} shows that the four most neutron-rich bins with $Y_e\leq 0.45$, are only sampled by three particles in the inline set. These particles are the main contributors to the neutron-rich isotopes between Zn and Zr. Among the backward reconstructed particles  only a single particle freezes out in this range of $Y_e$ and thus the yields, which are already low with the inline tracked particles, are even lower for the backward reconstructed particles. 
\begin{figure}
    \centering
    \includegraphics[width=\linewidth]{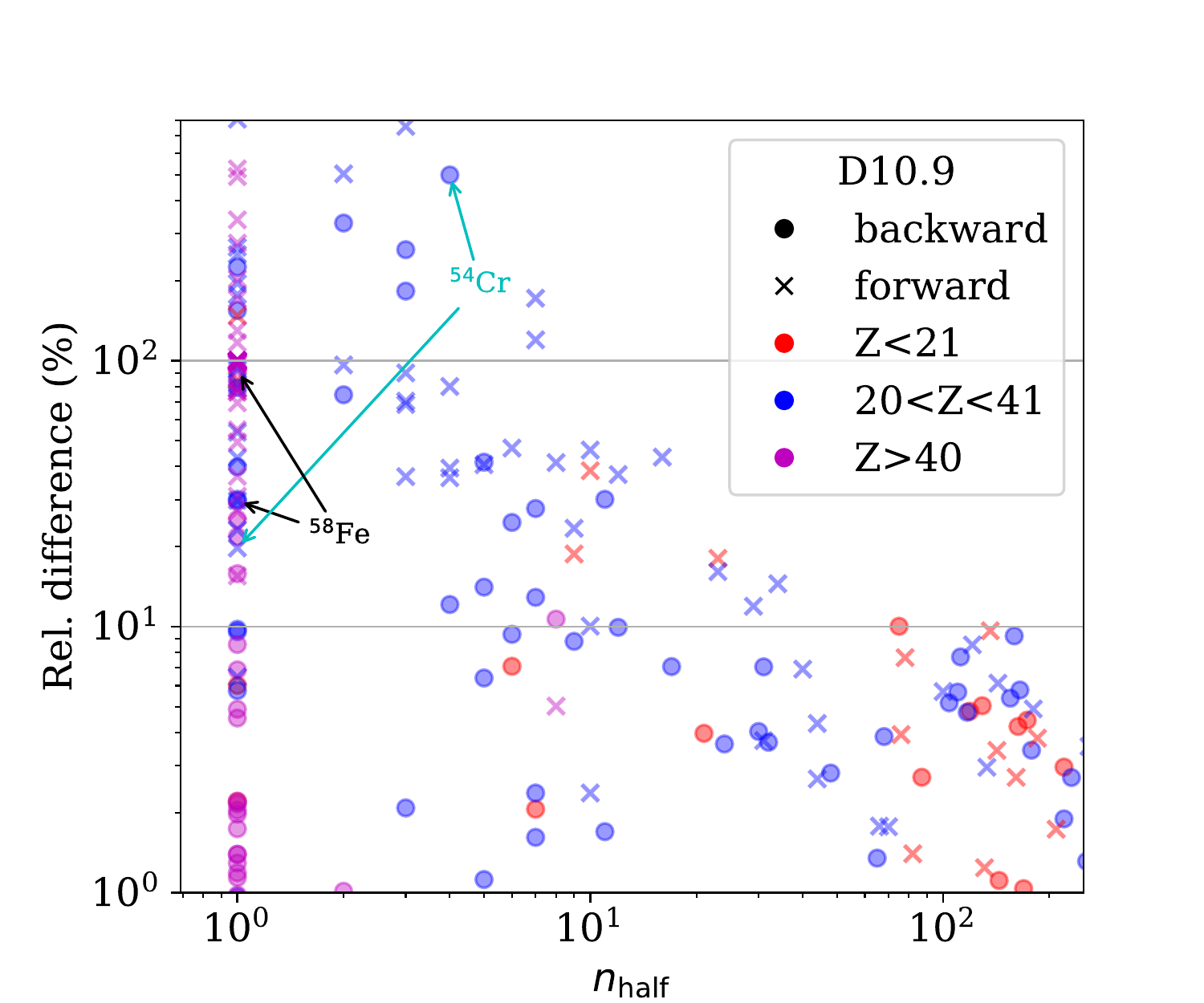}
    \caption{Same as Figure \ref{fig:nhalf_D96} but for model D10.9. The total number of particles is larger in this case and the maximum deviations smaller.  }
    \label{fig:nhalf_D109}
\end{figure}
Figure \ref{fig:ye_histo_d109} also shows that the proton-rich tail of the distribution, which mostly corresponds to late-time ejecta, is well reproduced by the backward reconstruction. In these wind-like ejecta, a weak $\nu \rm{p}$ process contributes to the yields of isotopes with $A>100$ up to Ag and Cd. The backward integration reproduces these yields remarkably well despite the fact that only very few particles sample these conditions. 
This emphasizes the advantage of the backward reconstruction for this case, because these ejecta only freeze out from NSE near the end of the simulation as they are ejected from the convective region, thus the backward integration can achieve an accurate reproduction of the freeze-out conditions without the need to track the particles through any complex fluid motions, leading to an overall very good agreement of the integrated nucleosynthesis yields.
To test the potential impact of the integration method, we also performed post-processing reaction network calculations with tracer particles using $N_{\rm{sub}}=10$ in the backward integration of the equations of motion. The final nucleosynthesis yields agree with the $N_{\rm{sub}}=40$ case to within a few percent for most isotopes, with the exception of the rarest isotopes, that show deviations of up to \percent{10}. 

Differences are larger for the forward reconstructed particles, as shown by the ratios in the lower panel of Figure \ref{fig:pf_comparison_d109}. 
However, for most isotopes differences remain smaller than a factor 2.
Again, the differences can be explained very well by the $Y_e$ distribution shown in Figure \ref{fig:ye_histo_d109}. 
Differences in \iso{46}{Ca}, \iso{50}{Ti} and \iso{58}{Fe} reflect the differences in the neutron-rich tail of the distribution, where the forward reconstruction is missing some particles in the $Y_e=0.455-0.465$ bin and has two additional particles with $Y_e<0.45$. 
The proton-rich tail of the $Y_e$~distribution is less populated with the forward reconstructed particles. Only 5 particle freeze out with $Y_e\geq 0.6$, but there is one single particle sampling a plume of material with very high $Y_e$ of 0.73. This plume appears very late in the simulation and is sampled by neither the inline nor the backward integrated particles. 
Since this particle represents conditions that are very different from all other particles, its impact on the overall yields is quite significant and it is responsible for the large production of \iso{96}{Mo}, Cd and Pd isotopes compared to the inline tracers. Note, however, that even though the differences of the yields of these isotopes are up to a factor 100, the production factors remain small compared to the much larger production factor for \iso{90}{Zr} which dominates the yield. 
To illustrate the impact of this proton-rich particle, the bottom panel of Figure \ref{fig:pf_comparison_d109} also shows the results when this particle is omitted as a thin, shaded line and small crosses. This clearly shows that the excess for $A>100$ is entirely due to this particle and when it is removed, the ratios are less than unity as a consequence of the 
general trend of the forward reconstruction to miss some of the most proton-rich ejecta.  

To illustrate the role of sampling and the number of tracer particles, Figure \ref{fig:nhalf_D109} shows the relative differences between the inline tracers and the reconstructed ones as a function of the value of $n_{\rm{half}}$, which is the minimum number of particles representing half of the yield of a given isotope, as in in Figure \ref{fig:nhalf_D96}. 
Since the total number of tracer particles is much larger in model D10.9, the scale for $n_{\rm{half}}$ is extended in Figure \ref{fig:nhalf_D109} compared to Figure \ref{fig:nhalf_D96}.
The neutron-rich isotopes \iso{58}{Fe} and \iso{54}{Cr} are highlighted and their production is dominated by a single or a few particles, leading to large differences. Similar to the case of D9.6, we find that the relative differences for isotopes that are represented by more than 20~tracer particles, are at most \percent{10}. This gives the scale for the uncertainties to expect from trace particle nucleosynthesis yields for isotopes that are well sampled. 

Overall, we find for model D10.9 that the backward reconstruction gives a better agreement for the integrated yields with respect to the inline tracked particles than the forward reconstruction, similar to our findings for model D9.6.
Due to the larger number of particles in model D10.9, however, the deviations tend to be smaller than what we found for D9.6. The ejecta from model D10.9 also do not have a strong neutron-rich component, which reduces the range of possible nucleosynthesis products. On the other hand, the nucleosynthesis of model D10.9 is complicated by the late-time ejection of very proton-rich material that is only sampled by the forward reconstructed particles, which probably overestimate its contribution to the integrated yield.

\subsection{Impact of the time resolution}

\begin{figure*}
    \centering
    \includegraphics[width=\linewidth]{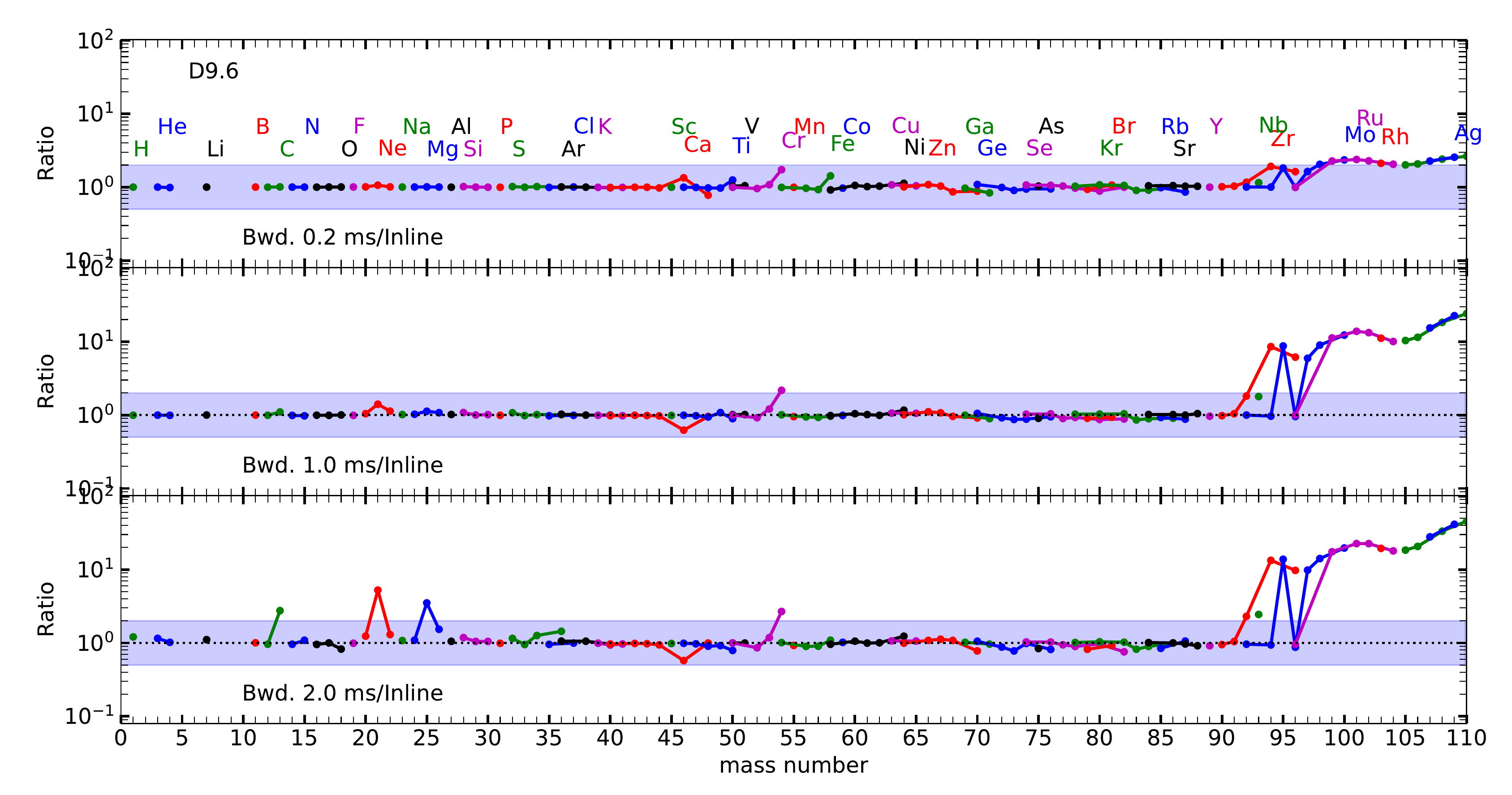}
    \caption{Ratios calculated as in Figure \ref{fig:pf_comparison} for model D9.6, but with the reconstruction based on increased snapshot intervals. The top panel correspond to our default value of 0.2 ms, which is the same as in Figure \ref{fig:pf_comparison}. The middle and bottom panel show the results with increased intervals of 1~ms and 2~ms, respectively.}
    \label{fig:interval}
\end{figure*}

The main limitation of tracer particle reconstruction is the limited
time-resolution of the available simulation data. 
While we cannot simply improve the time resolution, we can compare the results based on the 0.2~ms interval to 
results based on particles reconstructed with longer time steps of 1~ms and 2~ms by skipping some of the simulation snapshots, effectively ignoring part of the information that is available, in order to mimic a situation in which simulation snapshots are less frequent. 
Note that we keep the number of substeps, $N_{\rm{substep}}=40$, effectively increasing the integration timestep.
We focus here on model D9.6, which is computationally less expensive because of the smaller number of particles.
Figure \ref{fig:interval} shows the ratios of the results with 
tracer particles extracted with different snapshot intervals with the backward reconstruction for model D9.6. The top panel corresponds to the default interval with the results discussed above. Comparing to the middle panel, which shows the results with an interval of 1~ms, it is clear that the deviations from the inline tracked particles increase. First,
the already noticeable deviations of isotopes such as \iso{46,48}{Ca},\iso{50}{Ti} , \iso{54}{Cr}, \iso{58}{Fe}, that have already been discussed in Section~\ref{sec:nucleosynthesis}, increase further. The deviation of the yields for isotopes with A$\,>\!100$, which are only produced with very low abundance by a few particles, increases drastically, but their production factors are very small as they are produced by individual tracer particles at the edges of the $Y_e$ distribution. 
For most of the isotopes, the deviations remain level of a few \% at most. In Figure \ref{fig:timescales}, we showed that the dynamical timescales increase very steeply with the initial radius and as a result, even a relatively large change of the snapshot interval does not drastically change the number of particles that cannot be accurately reproduced. Therefore, we can expect a relatively mild dependence of the accuracy of the reproduction on the snapshot interval. 
With an interval of 2~ms, shown in the bottom panel of Figure \ref{fig:interval}, noticeable deviations for many common isotopes start to appear at the \percent{10} level. The isotopes \iso{13}{C}, \iso{21}{Ne} and \iso{25}{Mg} even show differences of more than a factor two.
This shows that, at this low resolution, the most common conditions that are represented by a larger number of particles start to be affected.
Based on this very limited parameter exploration, we suggest that a snapshot interval of 1~ms or better allows for reasonable nucleosynthesis results with backward reconstruction. Longer intervals, 2~ms or , should be avoided, because even the most common isotopes can be affected.

In the forward reconstruction, we find a similar increase of deviations as the snapshot interval is increased. Furthermore, the fate of individual particles depends substantially on the reconstruction. 
We find that the worsening the time resolution leads to a systematic reductions in the number of ejected tracer particles and thus of the ejecta mass. 
We find 598, 587 and 546 ejected particle for the reconstruction with  0.2~ms, 1~ms and 2~ms time steps respectively, compared to 603 ejected particles with the inline tracking.
This indicates that following the complex motion of neutrino heated matter
requires a high time resolution. 
Once a particle is close to the PNS, it is more likely to remain bound, if the time resolution is too low to capture the detailed evolution. 
With the forward reconstruction, 
large differences also quickly appear for the important radioactive isotope \iso{56}{Ni}. With the inline tracers, we obtain $5.18 \times 10^{-3}$~\msun\ of \iso{56}{Ni} and we find very similar values of around $5.33 \times 10^{-3}$~\msun\ for the forward reconstructed particles with time steps of 0.2~ms and $5.29 \times 10^{-3}$~\msun\ 1~ms. With 2~ms, however, the \iso{56}{Ni} yield is reduced to $3.49 \times 10^{-3}$~\msun. With the backward reconstruction in 2~ms intervals, we find $4.65 \times 10^{-3}$~\msun\ of \iso{56}{Ni}, i.e., the inline result is still reproduced to within \percent{10}.
For forward reconstruction, the snapshot interval seems to be even more important.  
We therefore suggest not to use intervals of more than 1~ms for any nucleosynthesis post-processing.
Even for inline tracers, \cite{Harris.Hix.ea:2017} found that using snapshots at intervals greater than 1~ms resulted in appreciable nucleosynthesis differences due simply to poor sampling of the temperature history.

\section{Conclusions}
\label{sec:conclusions}
We have compared, for the first time, nucleosynthesis calculations with inline tracked tracer particles directly from CCSN simulations to the results based on trajectories of tracer particles obtained in a post-processing step from snapshots of the same simulations. The reconstruction of particles is frequently used in the literature, in cases where the simulation did not track any tracer particles. We focus on the comparison between two method of tracer particle reconstruction: forward and backward integration in time, looking at two different axisymmetric simulations based on zero-metallicity progenitor models with masses of  9.6 and 10.9~\msun\ respectively. We use the initial and final positions of the inline particles for the reconstruction methods to allow a one-to-one comparison of tracer particles. In the absence of inline particles, different methods for the positioning of the particles need to be employed that may improve the accuracy of the reconstruction.
Compared to our approach to forward reconstruction, we find that the backward reconstruction has the advantage that it tends to reproduce the freeze-out conditions, i.e., the conditions when a particle cools down below about 8~GK, more accurately because it does not need to follow the particles through the complicate fluid motions induced by neutrino-heating and convection. The backward reconstruction, thus, results in integrated yields that are in better agreement with the inline-tracked particles than forward reconstructed particles with our approach. Note, however, that this might not apply to methods of forward reconstruction that avoid tracking particles through the neutrino-heating phase, e.g. by starting the forward integration at later times to gain the same advantage as backward tracing. 
We also show that the reconstruction of the initial position of a given particle in the progenitor, are not very reliable for any particles that get involved in convective and turbulent motion induced by neutrino heating because the fluid timescales are much shorter than the sampling interval. 
Furthermore, we find that the forward reconstruction tends to underestimate the amount of proton-rich material that is ejected at late times while the backward reconstruction is more prone to underestimate early, neutron-rich ejecta.
In general, however, the yields of both methods agree well for all isotopes that are produced by a sufficiently large number of particles (typically more than 20). Significant deviations appear only for isotopes that are represented by a smaller number of particles, emphasizing the need for very high mass resolution when only reconstructed tracers are available. We have also investigated different time intervals of 0.2, 1 and 2~ms between simulation snapshots for model D9.6 and find only a mild dependence of the level of agreement of the integrated yield on the snapshot interval. The agreement for the most common isotopes, however, deteriorates noticeably with an interval of 2~ms, affecting even more common isotopes, such as \iso{21}{Ne} and \iso{25}{Mg}, as well as \iso{56}{Ni} in the case of forward reconstruction.  With our setup, a 1~ms snapshot interval allows reproduction of the inline results of the important \iso{56}{Ni} and \iso{44}{Ti} yields to within \percent{10}. Hence, we recommend the use of backward integration with snapshots intervals of 1~ms or less for nucleosynthesis post-processing to obtain integrated yields, when inline tracers are unavailable.
While backward reconstruction thus seems to be a reasonable approach for nucleosynthesis purposes, it is not necessarily suitable to understand the fluid motions, due to the difficulties to reconstruct the initial position for particles that pass through the neutrino heating region. 
The enhanced utility of backward tracers is predicated on the ability of NSE to erase the need for detailed thermodynamics histories prior to freeze-out for particles that experience neutrino-driven convection.

In our analysis we assume that the inline tracked tracer particles provide the fiducial case. 
Another important comparison, however, can be done if a larger reaction network can be incorporated in the simulation itself, and compared to the post-processing, as in \citet{Navo.Reichert.ea:2022} and \citet{Harris.Hix.ea:2017}. 
Such a comparison allows the evaluation of the effects of the mass resolution of the tracer particles as well as the impact of mixing, which is possible between grid cells, but completely neglected by the tracer particle method.  
Our results are based on two axisymmetric models and will need to be confirmed with additional simulations, especially full 3D models, to verify our findings.
 
\acknowledgments
This research was supported by the U.S. Department of Energy Offices of Nuclear Physics and Advanced Scientific Computing Research;
the NASA Astrophysics Theory Program (grant NNH11AQ72I); 
and the National Science Foundation PetaApps Program (grants  OCI-0749242, OCI-0749204, and OCI-0749248), Nuclear Theory Program (PHY-1913531, PHY-1516197) and Stellar Astronomy and Astrophysics program (AST-0653376). AS acknowledges funding by the European Union’s
Framework Programme for Research and Innovation Horizon 2020 under Marie Sklodowska-Curie grant agreement No. 101065891. 
This research used resources of the National Energy Research Scientific Computing Center (NERSC), a U.S. Department of Energy Office of Science User Facility located at Lawrence Berkeley National Laboratory, operated under Contract No. DE-AC02-05CH11231.
This research used resources of the Oak Ridge Leadership Computing Facility at Oak Ridge National Laboratory.
Research at Oak Ridge National Laboratory is supported under contract DE-AC05-00OR22725  from the Office of Science of the U.S. Department of Energy to UT-Battelle, LLC. \\

\facility{NERSC,OLCF}
\newpage
\bibliography{bib,ccsn,math,astro}

\end{document}